\RequirePackage{fix-cm}
\documentclass{svjour3}                     
\smartqed  
\usepackage{comment}
\usepackage{color}
\usepackage[titletoc,title]{appendix}
\usepackage{amsmath}
\usepackage{amssymb}

\usepackage[authoryear]{natbib}
\usepackage{graphicx}
\usepackage{array}
\setcitestyle{notesep={; },round,aysep={},yysep={;}}
\usepackage[colorlinks,linkcolor=blue,anchorcolor=red,citecolor=green]{hyperref}
 \usepackage{mathptmx}      
\begin{document}

\title{Secular Resonances Between Bodies on Close Orbits:}


\subtitle{A Case Study of the Himalia Prograde Group of Jovian Irregular Satellites}

\titlerunning{Secular Resonances Between Bodies on Close Orbits}        

\author{Daohai Li         \and
        Apostolos A. Christou 
}


\institute{D. Li \at
              Armagh Observatory, College Hill, Armagh, BT61 9DG, Northern Ireland, UK \\
              School of Mathematics and Physics, Queen's University Belfast, University Road, Belfast, BT7 1NN, Northern Ireland, UK\\
              Tel.: +44 (0)28 3752 2928\\
              Fax: +44 (0)28 3752 7174\\
              \email{lidaohai@gmail.com; dli@arm.ac.uk}           
           \and
            A. A. Christou \at
              Armagh Observatory, College Hill, Armagh BT61 9DG, Northern Ireland, UK 
}

\date{Received: date / Accepted: date}

\maketitle

\begin{abstract}
 The gravitational interaction between two objects on similar orbits can effect noticeable changes in the orbital evolution even if the ratio of their masses to that of the central body is vanishingly small. \citet{Christou2005} observed an occasional resonant lock in the differential node $\Delta \Omega$ between two members in the Himalia irregular satellite group of Jupiter in the $N$-body simulations (corresponding mass ratio $\sim 10^{-9}$). Using a semianalytical approach, we have reproduced this phenomenon. We also demonstrate the existence of two additional types of resonance, involving angle differences $\Delta\omega$ and $\Delta (\Omega+\varpi)$ between two group members. These resonances cause secular oscillations in eccentricity and/or inclination on timescales $\sim$ 1 Myr. We locate these resonances in $(a,e,i)$ space and analyse their topological structure. In subsequent $N$-body simulations, we confirm these three resonances and find a fourth one involving $\Delta \varpi$. In addition, we study the occurrence rates and the stability of the four resonances from a statistical perspective by integrating 1000 test particles for 100 Myr. We find $\sim 10-30$ librators for each of the resonances. Particularly, the nodal resonance found by \citeauthor{Christou2005} is the most stable: 2 particles are observed to stay in libration for the entire integration.

\end{abstract}


\section{Introduction}
\label{sec-intro}

All planets in the solar system except Mercury and Venus have satellites \citep[see, e.g.,][]{Murray1999}. Satellites are subject to many types of perturbations, for instance, solar forcing and the oblateness of the host planet.

The Sun's gravity induces a satellite's orbit to precess at a rate, \citep[see, for example,][]{Innanen1997}
\begin{equation}
\dot \varpi_\odot \approx{3 k^2 m_\odot \over { 4 n a_\odot^3}}\,,
\end{equation}
where $\varpi$ is the longitude of pericentre of the satellite, $k^2$ the gravitational constant; $m_\odot$ and $a_\odot$ are the solar mass and semimajor axis of the Sun's relative motion with respect to the planet; $n$ is the satellite's mean motion and $n=\sqrt{k^2 m / a^3}$ (where $m$ is the mass of the host planet and $a$ the semimajor axis of the satellite). On the other hand, the oblateness of the host planet also causes the orbit to precess at the rate \citep[e.g.,][]{Roy1978}
\begin{equation}
\dot \varpi_\mathrm{J_2} \approx {3 \over 2} n J_2 {R^2 \over a^2}\,,
\end{equation}
where $J_2$ is a measure of the oblateness and $R$ is the equatorial radius of the planet. 
Equating the two, we find the critical semimajor axis
\begin{equation}
a_\mathrm{crit}=\left({2J_2R^2a_\odot^3 m \over m_\odot}\right)^{1/5}\,.
\end{equation}
For a satellite with $a$ below this value, its orbital evolution is mainly controlled by the planetary oblateness, while one with $a$ above this value is governed by solar perturbations. Based on this, \citet{Burns1986} defined irregular satellites as those whose orbital evolution is dominated by the Sun, i.e., with $a>a_\mathrm{crit}$ \citep[see also,][]{Goldreich1966}. Thus inherently, this definition mostly places constraints on the average distance from the satellite to the host planet. Apart from the size, their orbits are often highly eccentric and inclined compared to the regular satellites.

The eccentricity and inclination of the irregulars are generally high and their distribution in $[0,1)$ and $[0^\circ,180^\circ)$ is not uniform \citep[see, e.g.,][]{Jewitt2007}. For example, the inclinations of them are far from $90^\circ$. This feature can be explained by Kozai-Lidov dynamics \citep{Lidov1962, Kozai1962} that involves the secular perturbation from the Sun. Under solar forcing, the eccentricity and inclination of the satellite are coupled in such a way that the vertical angular momentum $H\propto\sqrt{1-e^2} \, \cos{i}$ is conserved while the eccentricity $e$ and inclination $i$ of the satellite may experience large-amplitude oscillations, especially when $i$ is high. This mechanism causes the absence of satellites with inclinations around $90^\circ$ \citep{Carruba2002}. We will discuss this effect in detail in Sect. \ref{semi-sec-total}.

Most irregular satellites reside on retrograde, rather than prograde, orbits; this relates to the so-called evection phenomenon \citep{Yokoyama2008, Frouard2010}. Evection is linked to the mean motion of the Sun. If a satellite is too far from the planet, it  will be so perturbed by the Sun that its longitude of pericentre $\varpi$ precesses at a rate comparable to the solar mean motion. In this case, the angle $\psi=\varpi-\lambda_\odot$ (where $\lambda_\odot$ is the mean longitude of Sun) may librate; the angular momentum $G \propto \sqrt{1-e^2}$ may oscillate strongly and the eccentricity may grow from $\sim 0$ to 0.6 \citep{Frouard2010}. Thus the evection effect places constraints on the stability of irregular satellites and influences their orbital distribution. Furthermore, this effect is not symmetrical with respect to the inclination of $90^\circ$ because for prograde orbits, $\varpi=\Omega+\omega$ (where $\Omega$ is the longitude of the ascending node and $\omega$ the argument of pericentre) while for retrograde orbits, $\varpi$ is the difference of the two angles. As a result, the eccentricity of a prograde orbit is excited more efficiently than a retrograde one, providing a possible mechanism to explain why the retrograde regime harbours more irregular satellites \citep{Nesvorny2003}.

A prominent feature of the irregular satellite population is that they form groups or families \citep{Nesvorny2003, Sheppard2003}. Members of the same family have similar orbital elements. Families are thought to be the products of collisional evolution. \citet{Nesvorny2003} used the Gauss equations to infer the degree of dispersion within different groups of irregular satellites. They found that the Himalia family is more widely dispersed than expected for a group of collisional fragments. \citet{Beauge2006} and \citet{Beauge2007} developed a high order analytical method and used it to calculate accurate precession rates of irregular satellites and to study the dynamical structure of the Kozai resonance. Based on this method, they determined the locations of secular resonances and compared them with the positions of known irregular satellites; also, a new family around Pasiphae was identified. \citet{Hinse2010} applied the MEGNO technique to jovian irregular satellites and identified chaotic and quasi-periodic regions; they also found some high-order mean motion resonances. Using similar methods, \citet{Frouard2011} reported a number of resonances related to the Great Inequality and revealed the chaotic diffusion in different irregular satellite groups.

All of the above work has assumed the irregular satellites to be massless. However, in a manner similar to asteroid families \citep[e.g.,][]{Carruba2003} a massive member of a satellite family could exert strong perturbations on the smaller members and scatter their orbits; \citet{Christou2005} found that, gravitational scattering by Himalia operating over the age of the solar system could be responsible for the large velocity dispersion observed by \citet{Nesvorny2003}. \citeauthor{Christou2005} used this effect to place constraints on the mass of Himalia. He also observed a transient resonant lock between the nodes of Himalia and Lysithea, another satellite in the family. Thus we suppose a massive member could have significant effects on the orbital evolution of the group and this phenomenon needs further study. Here, as a case study, we focus our attention on the Himalia group of Jupiter and particularly on the resonances identified by \citeauthor{Christou2005}.

The paper is organised as follows: in Sect. \ref{nodal-intro-sec}, we introduce the Himalia group, the phenomenon of ``nodal libration'' and the model used to study it. Then in Sect. \ref{semi-sec-total}, we describe our semianalytical approach to model the nodal libration and introduce other resonant phenomena. The results of $N$-body simulations are presented in Sect. \ref{sec-n-body}. Finally, Sect. \ref{sec-discussion} is devoted to the conclusions and discussions.


\section{The Himalia group, a nodal resonance and our model}
\label{nodal-intro-sec}
The Himalia group is a prograde jovian irregular satellite group with five members: Himalia, Elara, Lysithea, Leda and Dia. Himalia is the largest member that carries the group name; it is at least 8 times more massive than any of the other four known members\footnote{Dia was discovered in 2000 and then lost. It was recently recovered by S. S. Sheppard; see \url{http://dtm.carnegiescience.edu/news/long-lost-moon-jupiter-found}. For the mass estimates of the members, see \url{http://ssd.jpl.nasa.gov/?sat_phys_par}. For orbital elements, see \url{http://ssd.jpl.nasa.gov/?sat_elem}. Parameters of other satellites from the latter two links will be used in Sect. \ref{sec-discussion}.}. \citet{Emelyanov2005} studied the perturbation of Himalia on other satellites and estimated the mass to be $k^2 m=(0.28 \pm 0.04)$  km$^3$/s$^2$, consistent with the estimate of \citet{Christou2005} based on its scattering effect. We adopt this mass estimate throughout the paper. The mean semimajor axis, eccentricity and inclination of the group are 0.078 AU, 0.18 and 28$^\circ$, respectively.

In \cite{Christou2005}, where the main purpose was to study the influence of gravitational scattering within the Himalia group, the effects of the Sun, Saturn and jovian oblateness (corrected for the additional mass of Galilean satellites) were considered; Himalia and Elara were assumed to be massive and other family members were massless. In such a configuration, it was found that the difference between the longitudes of the nodes of Lysithea and Himalia $\Delta \Omega=\Omega_\mathrm{L}-\Omega_\mathrm{H}$ could occasionally librate around $\pi$ on a timescale of 1 Myr with the amplitude of the oscillation in inclination being ${}\sim 0.1 ^\circ$ (see figs. 4 and 5 of that work). \citeauthor{Christou2005} also reported that, on entering and leaving the libration, significant changes in the inclination can occur. This provides a fast channel for the evolution of inclination and may contribute to the large relative velocity dispersion in the Himalia group. We refer to this phenomenon as ``nodal resonance'' or ``nodal libration'' hereafter.
 
\begin{figure}[h]
\begin{center}
	\includegraphics[width=0.7\textwidth]{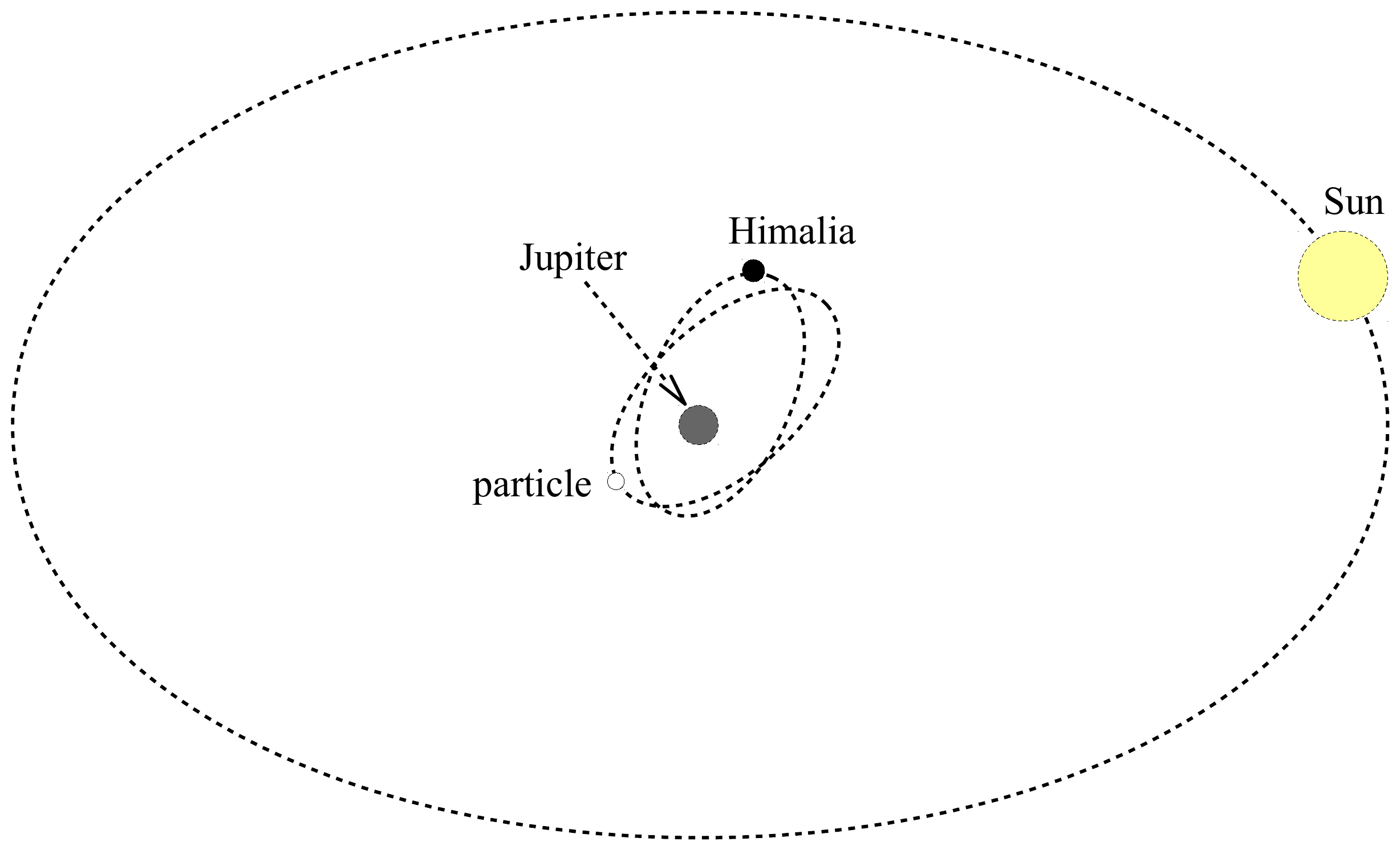}
\caption{Illustration of the system that consists of Jupiter, the Sun, Himalia and a particle}
\label{model-illustration}
\end{center}
\end{figure}

Here, our goal is to specifically study this dynamical phenomenon. To isolate the relevant dynamics, we work within a restricted four-body problem model containing Jupiter, the Sun, a massive Himalia and a massless test particle; see Fig \ref{model-illustration}. Here, all bodies revolve around Jupiter (in the relative sense). The Sun is assumed to be on a circular orbit.

In the following sections, we tackle this four-body problem with both semianalytical and fully numerical methods.

\section{Semianalytical approach}
\label{semi-sec-total}

Our strategy is to split the four-body problem of Jupiter-Sun-Himalia-particle into two restricted three-body problems that can be tackled separately using perturbation analysis. The first includes Jupiter, the Sun and Himalia. Since the mass of Himalia is vanishingly small compared with those of Jupiter and the Sun, it is assumed to be also massless in this part. Thus Himalia and the particle can be studied using the same technique. The second three-body problem is comprised of Jupiter, massive Himalia and the massless particle. We describe in detail the methods dealing with the two restricted three-body problems in the following subsections. Then we explain how we combine the two.

In this paper, we use a normalised unit system in which the Solar mass is 1; jovian semimajor axis is 1; gravitational constant is 1; thus the orbital period of Jupiter is approximately $2 \pi$; in practice, we have used $2 \pi$ in time conversion for the $N$-body simulations described in Sect. \ref{sec-n-body}.


\subsection{Kozai dynamics in the Jupiter-Sun-satellite restricted three-body problem}
\label{sec-solar}

In the first instance, solar perturbations dominate the evolution of Himalia and the particle. A substantial body of literature has dealt with such ``hierarchical'' planet-Sun-satellite problems and has discussed the Kozai-Lidov mechanism \citep{Lidov1962,Kozai1962}. There, the Hamiltonian of the system is expanded in terms of the ratio of semimajor axis of the satellite to that of the Sun (we denote this ratio as $\alpha=a/a_\odot$ hereafter) and only the leading order terms (of order $\alpha^2$; the ``quadrupole'' Hamiltonian) are retained. Actually, in our simple model, the Sun has zero eccentricity, causing the octupole Hamiltonian (of order $\alpha^3$) to be zero \cite[see, e.g.,][]{Krymolowski1999, Naoz2013}. Then the mean anomalies of the Sun and Himalia are eliminated. Owing to a ``happy coincidence'' \citep{Lidov1976}, the quadrupole secular Hamiltonian is integrable. Specifically, this Hamiltonian is
\begin{equation}
\label{quad_initial}
F={k^2 m_\odot a^2\over 16 a_\odot^3} \left[ \left(-15e^2\cos \left(2 \omega\right)+9e^2+6\right)\cos ^2 i+15e^2\cos \left(2 \omega\right)-3e^2-2\right],
\end{equation}
in which $k^2$ is the gravitational constant and $m_\odot$ the mass of the Sun; $a$ and $a_\odot$ are semimajor axes of the satellite and the Sun; $e$, $i$ and $\omega$ are the satellite's orbital eccentricity, inclination and argument of pericentre, respectively \citep{Kozai1962, Naoz2013}; this Hamiltonian and its simplified forms will be referred to as the ``Kozai Hamiltonian'' hereafter. Apparently, the Hamiltonian $F$ and semimajor axis $a$ are conserved. As the longitude of the ascending node $\Omega$ does not appear in Eq. \eqref{quad_initial}, the vertical component of the angular momentum $H\propto \sqrt{1-e^2} \,\cos i$ is constant as well. For a dynamical system governed by this Hamiltonian, $\omega$ may librate around $\pm \pi/2$ while the angular momentum $G\propto \sqrt{1-e^2}$ may experience large amplitude oscillation, giving rise to the Kozai resonance. All four giant planets have such satellites \citep{Nicholson2008}. The libration in $\omega$ requires the inclination to exceed $\approx 40^\circ$; this is not the situation for the Himalia group. Generally, such evolution under solar perturbation is referred to as a ``Kozai cycle''. In Kozai theory, the reference plane is the orbital plane of  Jupiter; this frame is used throughout the paper.

With this Hamiltonian, we can obtain the expressions for the nodal and apsidal precession rates of the satellite. The Hamiltonian will be further simplified in Subsect. \ref{sec-analytical}.


\subsection{Satellite interaction in the Jupiter-Himalia-particle restricted three-body problem}
\label{sec-coorbital}

Here, we consider the subsystem comprised of Jupiter, Himalia and a massless test particle.

Since the orbits of the two satellites are close and may cross, the traditional expansion of the disturbing function \cite[e.g.,][]{Murray1999} is not applicable here. To tackle this, we resort to treating the problem under the framework of coorbital theory. The crucial differences between coorbital theory and the conventional treatment lie in the expansion in the disturbing function and in the elimination of the mean anomalies. In coorbital theory, no expansion is made with respect to the ratio of semimajor axes. The mean anomalies of the two orbits are eliminated in a combined manner so that the difference in the mean longitudes $\Delta \lambda$ remains. This enables the study of the variation of the relative semimajor axis $\Delta a$ \cite[see, for example,][]{Message1966, Morais1999}. When studying the secular evolution of the system,  $\Delta \lambda$ is eliminated.

We follow \cite{Henon1986} \citep[see also,][]{Namouni1999} and use the relative quantities, e.g., the relative vector eccentricity $\mathbf{e}_\mathrm{r} =\mathbf{e}_\mathrm{P}-\mathbf{e}_\mathrm{H}=(e_\mathrm{P} \cos \varpi_\mathrm{P}-e_\mathrm{H} \cos \varpi_\mathrm{H}, e_\mathrm{P} \sin \varpi_\mathrm{P}-e_\mathrm{H} \sin \varpi_\mathrm{H})$ (where $\varpi$ is the longitude of pericentre) of the particle relative to Himalia.

We assume that Himalia and the particle are on similar orbits (i.e., with similar semimajor axes, eccentricities and inclinations), that their eccentricities and inclinations are much smaller than unity and that their total mass is far smaller than that of the central body, Jupiter. The semimajor axes, eccentricities and inclinations of the members of the group are restricted to small ranges and the eccentricities are relatively small, satisfying the requirement of coorbital theory. However, the inclination of the Himalia group is about 0.5 rad. As the main purpose of this section is to describe the dynamical structure qualitatively but not to provide a precise quantitative description of the orbital evolution, we divide all inclinations by a factor of 3 in this section wherever coorbital theory is involved. In this way, the eccentricities and inclinations are all around 0.17, better suiting the expansion described below. We will briefly discuss what the situation is if we apply the original inclinations in Sect. \ref{sec-discussion}. In this setting, the motion of the two satellites can be expressed in a frame whose origin revolves around Jupiter on a circular orbit of semimajor axis $a_0$ ($a_0\in\left(\min(a_\mathrm{H},a_\mathrm{P}),\max(a_\mathrm{H},a_\mathrm{P})\right)$); the frame also rotates along its $z$-axis at an angular velocity of $n_0=\sqrt{k^2 (m_\mathrm{J}+m_\mathrm{H})/a_0^3}$ (where $m_\mathrm{J}$ and $m_\mathrm{H}$ are masses of Jupiter and Himalia, respectively; the particle is massless). Omitting higher order terms in $e$ and $i$ and assuming that the two satellites are not generally interacting, the Cartesian coordinates of Himalia in this rotating frame can be expressed as the following functions of time \citep{Henon1986,Namouni1999}
\begin{equation}
\label{co-motion}
\left\{
\begin{aligned}
    &x_\mathrm{H}=a_\mathrm{H} (h_\mathrm{H} \cos n_0 t +k_\mathrm{H} \sin n_0 t)+(a_\mathrm{H}-a_0)
     \\
    &y_\mathrm{H}=-2 a_\mathrm{H} (h_\mathrm{H} \sin n_0 t -k_\mathrm{H} \cos n_0 t) - {3 \over 2} (a_\mathrm{H}-a_0)  (n_0 t-{ \lambda_\mathrm{H}})
     \\
    &z_\mathrm{H}=a_\mathrm{H}(p_\mathrm{H} \cos n_0 t +q_\mathrm{H} \sin n_0 t) \, ,
\end{aligned}
\right.
\end{equation}
where $a_\mathrm{H}$, $h_\mathrm{H}=e_\mathrm{H} \cos{\varpi_\mathrm{H}}$, $k_\mathrm{H}=e_\mathrm{H} \sin{\varpi_\mathrm{H}}$, $p_\mathrm{H}=i_\mathrm{H} \cos{\Omega_\mathrm{H}}$, $q_\mathrm{H}=i_\mathrm{H} \sin{\Omega_\mathrm{H}}$ and $\lambda_\mathrm{H}$ are the regular orbital elements of Himalia; similar equations apply for the particle. From the linearity of these expressions, it can be seen that analogous expressions exist for the motion of the particle relative to Himalia and the relative $e$ and $i$ are constant.

On account of a massive Himalia, the relative eccentricity and inclination begin to evolve. It has been shown the secular evolution of relative $e$ and $i$ observe the following potential \citep{Luciani1995,Namouni1999}
\begin{equation}
\label{co-potential}
R={{ k^2 \mu(m_\mathrm{J}+m_\mathrm{H}) }\over{2 \pi a_0} } \log \Delta\, ,
\end{equation}
where when $e_\mathrm{r} \left| \cos \omega_\mathrm{r}\right| \geq \left| a_\mathrm{r}\right|$,
\begin{equation}
\label{co-potential-one}
\Delta=2 e_\mathrm{r} i_\mathrm{r} \left| \cos \omega_\mathrm{r}\right| +i_\mathrm{r}^2+ e_\mathrm{r}^2 \, ,
\end{equation}
and when $e_\mathrm{r} \left| \cos \omega_\mathrm{r}\right| < \left| a_\mathrm{r}\right|$,
\begin{equation}
\begin{aligned}
\Delta&=\sqrt{2} \left| a_\mathrm{r}\right|  \sqrt{\sqrt{4 e_\mathrm{r}^2  i_\mathrm{r}^2 \sin ^2\omega_\mathrm{r}+\left(-i_\mathrm{r}^2+e_\mathrm{r}^2-a_\mathrm{r}^2\right)^2}+i_\mathrm{r}^2-e_\mathrm{r}^2+a_\mathrm{r}^2}
\\
&+\sqrt{4 e_\mathrm{r}^2 i_\mathrm{r}^2 \sin ^2\omega_\mathrm{r }+\left(-i_\mathrm{r}^2+e_\mathrm{r}^2-a_\mathrm{r}^2\right)^2}+a_\mathrm{r}^2 \, .
\end{aligned}
\end{equation}
In the above expressions, $\mu=m_\mathrm{H}/m_\mathrm{J}$, $a_\mathrm{r}=(a_\mathrm{P}-a_\mathrm{H})/a_0$, $e_\mathrm{r}=\sqrt{h_\mathrm{r} ^2+ k_\mathrm{r} ^2}$, $i_\mathrm{r}=\sqrt{p_\mathrm{r} ^2+ q_r ^2}$, $\varpi_\mathrm{r}=\tan^{-1} (k_\mathrm{r}/h_\mathrm{r})$, $\Omega_\mathrm{r}=\tan^{-1} (q_\mathrm{r}/p_\mathrm{r})$, $\omega_\mathrm{r}=\varpi_\mathrm{r}-\Omega_\mathrm{r}$, $h_\mathrm{r}=h_\mathrm{P}-h_\mathrm{H}$, $k_\mathrm{r}=k_\mathrm{P}-k_\mathrm{H}$, $p_\mathrm{r}=p_\mathrm{P}-p_\mathrm{H}$ and $q_\mathrm{r}=q_\mathrm{P}-q_\mathrm{H}$ (the relative elements). The equations of motion are then \citep{Namouni1999}
\begin{equation}
\label{co-euqation-motion}
\begin{aligned}
\dot h_\mathrm{r}&= {1\over n_0}{{\partial R}\over{\partial k_\mathrm{r}}} ,\qquad  \dot k_\mathrm{r}=-{1\over n_0}{{\partial R}\over{\partial h_\mathrm{r}}} ,
\\
\dot p_\mathrm{r}&= {1\over n_0}{{\partial R}\over{\partial q_\mathrm{r}}} ,\qquad  \dot q_\mathrm{r}= -{1\over n_0}{{\partial R}\over{\partial p_\mathrm{r}}} .
\end{aligned}
\end{equation}

The coorbital secular potential \eqref{co-potential} causes variations in $e_r$ and $i_r$; it also gives rise to precession in $\varpi_r$ and $\Omega_r$ \citep{Namouni1999}. These effects, in turn, are reflected in the evolution of the particle's elements $(\varpi_\mathrm{P}, e_\mathrm{P})$ and $(\Omega_\mathrm{P},i_\mathrm{P})$.


\subsection{Secular dynamics of the Jupiter-Sun-Himalia-particle system}
\label{sec-analytical}

Having isolated the relevant dynamics in the two sub-problems discussed in Subsects. \ref{sec-solar} and \ref{sec-coorbital}, we now want to combine them. This is done so that both Himalia and the test particle experience solar perturbation through the Kozai Hamiltonian \eqref{quad_initial}; to this we add the gravitational influence on the particle from Himalia by applying \eqref{co-euqation-motion} to potential \eqref{co-potential} and adding the corresponding variation rates to the particle. Thus the equations of motion are
\begin{equation}
\label{both-equation-motion}
\dot \sigma_\mathrm{H}=f^\odot(\sigma_\mathrm{H})   \quad \mathrm{and} \quad \dot \sigma_\mathrm{P}=f^\odot(\sigma_\mathrm{P})+f^\mathrm{coorb}(\sigma_\mathrm{r})\,,
\end{equation}
where $\sigma$ is an arbitrary orbital element; $f^\odot$ refers to the solar perturbation and is a function of the elements of Himalia or the particle; $f^\mathrm{coorb}$ represents the contribution from the coorbital interaction potential and is a function of the relative elements. We now have an eight-dimensional system of $(\varpi_\mathrm{H}, e_\mathrm{H}; \varpi_\mathrm{P}, e_\mathrm{P}; \Omega_\mathrm{H}, i_\mathrm{H}; \Omega_\mathrm{P}, i_\mathrm{P})$.

The evolution of the particle is generally dominated by solar perturbations even when perturbations by both the Sun and Himalia are considered. We use a simple example to demonstrate this point. According to Eqs. \eqref{quad_initial} and \eqref{co-potential}, and for the nominal orbits of Himalia and Lysithea (as the particle), the timescales of precession in $\varpi_\mathrm{P}$ and $\Omega_\mathrm{P}$ induced by the coorbital potential are $\tau_\varpi ^{\mathrm{coorb}} \sim 2 \pi/\dot \varpi ^{\mathrm{coorb}}\approx 2.7 \times 10^7$ and $\tau_\Omega ^{\mathrm{coorb}} \sim 2 \pi/\dot \Omega ^{\mathrm{coorb}}\approx 2.8 \times 10^7$. For solar perturbations, the corresponding timescales are $\tau_\varpi ^{\mathrm{\odot}} \sim 2 \pi/\dot \varpi ^{\mathrm{\odot}}\approx 1.5 \times 10^2$ and $\tau_\Omega ^{\mathrm{\odot}} \sim 2 \pi/\dot \Omega ^{\mathrm{\odot}}\approx 1.4 \times 10^2$. Noting that, according to \citet{Christou2005}, the timescale of nodal libration is $\sim 10^6$, thus the effects of a Kozai cycle should be considered as ``fast'' and can be eliminated when tackling nodal libration.

Assuming that within a Kozai cycle, changes in eccentricity is small ($\Delta e \ll 1$) and applying some algebraic manipulations (see Appendix \ref{sec-kozai-derivation} for details), we have the following ``averaged'' Kozai Hamiltonian
\begin{equation}
\label{constant-kozai}
F_\mathrm{K}=-{k^2 m_\odot a^2\over 16 a_\odot^3}\left(2+3e^2\right)\left(1-3\cos ^2i\right),
\end{equation}
where $e$ and $i$ are the ``averaged'' quantities over a Kozai cycle compared to the original ones in Eq. \eqref{quad_initial}. With this Hamiltonian, the precession rates of $\Omega$ and $\varpi$ are constant. Rates of group members are close with similar $a$, $e$ and $i$.

\begin{figure}[h]
\begin{center}
	\includegraphics[width=0.8\textwidth]{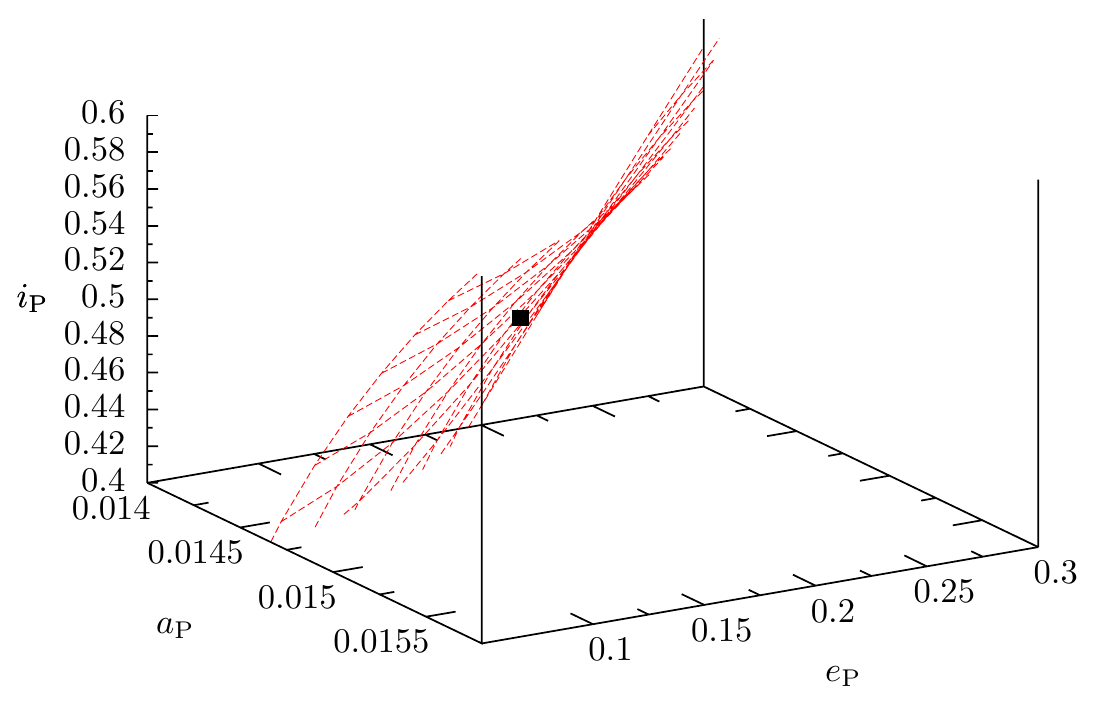}
\caption{The surface of equal precession rate in $\Omega$ in $(a,e,i)$ space. The location of the surface is calculated using the Kozai Hamiltonian \eqref{constant-kozai} without expansion in eccentricity or inclination. Points on the surface share the same precession rate as Himalia in $ \Omega$. The black square represents the location of Himalia}
\label{equal-precession-hh}
\end{center}
\end{figure}

As shown above, the influence of Himalia is several orders of magnitude smaller than that of the Sun. We suggest that, the libration in $\Delta \Omega$ is only possible where $\mathrm {d}\Delta \Omega / \mathrm{d} t$ is vanishingly small under solar perturbations. With the Hamiltonian \eqref{constant-kozai}, we define a surface in $(a,e,i)$ space where all points on it share the same nodal precession rate with Himalia: $\dot {\Omega} (a,e,i)=\dot {\Omega}|_\mathrm{H}$ (since $\dot {\Omega}$ is a function depending only on $a$, $e$ and $i$); we call this a Surface of Equal Precession Rate (SEPR) in $\Omega$. We show the SEPR in Fig. \ref{equal-precession-hh}; it is conceptually similar to those discussed in \citet{Williams1981} in the contact of secular resonances in the main asteroid belt. Note that, until now, we do not assume small eccentricity or inclination for the Kozai Hamiltonian; thus the SEPR applies to the real Himalia group, whose inclination is about 0.5 rad.

However, as the coorbital potential \eqref{co-potential} results from an expansion in eccentricity and inclination, combining it with the full expression for the Kozai potential \eqref{constant-kozai} seems superfluous. Hence, we also expand \eqref{constant-kozai} in $e$ and $i$, retaining terms up to order four to arrive at the following Kozai Hamiltonian
\begin{equation}
\label{orig-min-kozai}
{ F}=-{k^2 m_\odot a^2\over 16 a_\odot^3}\left(2i^{4}-9e^2i^2-6i^2+6e^2\right) \,.
\end{equation}
We then rewrite the Hamiltonian in normalised Poincar\'{e} conjugate variables, and again we adopt expansions in $e$ and $i$. The variables are
\begin{equation}
\bar g= -\varpi=-(\Omega+\omega), \bar G=e^2/2; \quad \bar h = -\Omega, \bar H=i^2/2.
\end{equation}
Under the new variable set, the Kozai Hamiltonian becomes
\begin{equation}
\label{min-kozai}
{ F}=-{{{k{m_\odot}a^{3/2}}}\over{4\sqrt{m_\mathrm{J}}\,{a_\odot}}^3}\left(2  \bar H^2-9  \bar G  \bar H-3  \bar H+3  \bar G\right) \,.
\end{equation}
Since now the variables are the normalised angular momenta, we introduce a factor of $k\sqrt{m_\mathrm{J} a}$ into the coefficient of \eqref{min-kozai} to preserve the Hamiltonian property. To better separate ``slow'' and ``fast'' angles, we apply the following canonical transformation:
\begin{equation}
\label{trans2slow}
\begin{aligned}
\bar g_\mathrm{f}&=&\bar g_\mathrm{P}, \quad \bar G_\mathrm{f}&=& \bar G_\mathrm{P}+\bar G_\mathrm{H};
\\
\bar g_\mathrm{s}&=&\bar g_\mathrm{P}-\bar g_\mathrm{H}, \quad \bar G_\mathrm{s}&=& -\bar G_\mathrm{H};
\\
\bar h_\mathrm{f}&=&\bar h_\mathrm{P}, \quad \bar H_\mathrm{f}&=& \bar H_\mathrm{P}+\bar H_\mathrm{H};
\\
\bar h_\mathrm{s}&=&\bar h_\mathrm{P}-\bar h_\mathrm{H}, \quad \bar H_\mathrm{s}&=& -\bar H_\mathrm{H}.
\end{aligned}
\end{equation}
On the right-hand side of the above expressions, $(\bar g,\,\bar G)$ and $(\bar h,\,\bar H)$ are conjugate normalised Poincar\'{e} variables with subscripts ``H'' and ``P'' referring to Himalia and the particle, respectively. On the left-hand side, angles (represented with lower-cased letters) with subscript ``f'' are faster compared to those with subscript ``s'', since the former are differences between, while the latter are identical to, the original angles; the conjugate momenta are represented with capital letters.

Thus the Kozai Hamiltonian of both Himalia and the particle assumes the following form
\begin{equation}
\label{slow-kozai}
\begin{aligned}
F_\mathrm{S}=C_\mathrm{P} &\left( 
2{\bar H_\mathrm{s}}^2+4{\bar H_\mathrm{f}}{\bar H_\mathrm{s}}-9{\bar G_\mathrm{s}}{\bar H_\mathrm{s}}-9{\bar G_\mathrm{f}}{\bar H_\mathrm{s}}-3{\bar H_\mathrm{s}}+2{\bar H_\mathrm{f}}^2  \right.
\\
&  \left. -9{\bar G_\mathrm{s}}{\bar H_\mathrm{f}}-9 {\bar G_\mathrm{f}}{\bar H_\mathrm{f}}-3{\bar H_\mathrm{f}}+3{\bar G_\mathrm{s}}+3{\bar G_\mathrm{f}}
 \right)
 \\
+C_\mathrm{H}&\left(
2{\bar H_\mathrm{s}}^2-9{\bar G_\mathrm{s}}{\bar H_\mathrm{s}}+3{\bar H_\mathrm{s}}-3{\bar G_\mathrm{s}}
\right) \,,
\end{aligned}
\end{equation}
with coefficients $C_\mathrm{P}=-{{k  { m_\odot} a_\mathrm{P}^{3/2}}/(4{ m_\mathrm{J}^{1/2}}{ a_\odot}^3})$ and $C_\mathrm{H}=-{{k { m_\odot} a_\mathrm{H}^{3/2} }/(4 { m_\mathrm{J}^{1/2}}{ a_\odot}^3})$, where $a_\mathrm{H}$ and $a_\mathrm{P}$ are the semimajor axes of Himalia and the test particle, respectively.

The coorbital potential \eqref{co-potential} is a branch function. Note that within the satellite group, the quantity $a_\mathrm{r}=(a_\mathrm{P}-a_\mathrm{H})/a_0$ is always small. Taking the orbital elements of Lysithea for example,  $a_\mathrm{r} \approx 0.02$ while $e_\mathrm{r}$ is of order $10^{-1}$. Thus the condition $e_{\mathrm{r}} \left| \cos \omega_{\mathrm{r}}\right| < a_{\mathrm{r}}$ is rarely satisfied and the potential form \eqref{co-potential-one} dominates. We omit the other form and presume that the potential only takes this form.

Again, we emphasise that, what matters is the difference between the nodes of the particle and Himalia. Hence, any effects that operate on timescales comparable to $2 \pi/ \dot\Omega$ are considered ``fast'' (cf. transformation \eqref{trans2slow}) and we need to eliminate them. Since we have assumed constant eccentricity and inclination in the Kozai theory \eqref{constant-kozai} and \eqref{min-kozai}, there is nothing to remove in the Kozai Hamiltonian. However, the short timescale effects in the coorbital potential \eqref{co-potential} need elimination. We proceed as follows. First, we substitute the variables in the coorbital potential to express it in the same variables as the Kozai Hamiltonian \eqref{slow-kozai}. Noting that the fast angles are controlled primarily by the Kozai potential, they precess evenly with time (cf. Eq. \eqref{slow-kozai}). Hence we perform two finite integrations and arrive at the ``slow'' coorbital potential\footnote{Though the form appears asymmetric in $e_\mathrm{r}$ and $i_\mathrm{r}$, exchanging $e_\mathrm{r}$ and $i_\mathrm{r}$ will not alter the value.}:
\begin{equation}
\label{slow-coorbital}
\begin{aligned}
R_\mathrm{S}=&{1\over{4 \pi^2}}\int_0^{2 \pi} \mathrm{d} \bar{g}_\mathrm{f} \int_0^{2 \pi} R(\bar g_\mathrm{f}, \bar G_\mathrm{f}; \bar h_\mathrm{f}, \bar H_\mathrm{f}; \bar g_\mathrm{s}, \bar G_\mathrm{s}; \bar h_\mathrm{s}, \bar H_\mathrm{s}) \, \mathrm{d} \bar{h}_\mathrm{f}
\\
=&{\mu n_0\over \pi^2}\left\{\Im\left[ \mathrm{Li}_2({{\rm i} i_\mathrm{r} \over e_\mathrm{r}}) - \mathrm{Li}_2(-{{\rm i} i_\mathrm{r} \over e_\mathrm{r}})  \right] + \log e_\mathrm{r} \right\}\,.
\end{aligned}
\end{equation}
Here, the variables have the same meaning as in Eqs. \eqref{co-potential} and \eqref{co-potential-one}; ${\rm i}$ is the unit imaginary number; $\Im$ represents the imaginary part; $\mathrm{Li}$ is the polylogarithm function defined as \citep[e.g.,][]{Weisstein1999}
\begin{equation}
\mathrm{Li}_n (z)=\sum_{k=1}^\infty {z^k\over k^n}
\end{equation}
where $z$ is a complex number and $|z|<1$.

We have now reached a system devoid of ``fast'' effects, i.e., those operating on timescales similar to a Kozai cycle. Note that the coorbital potential is {\it not} a Hamiltonian in these variables (we show this in the Appendix \ref{sec-non-canonical}). However, we can still write down the equations of motion of the system as
\begin{equation}
\label{equation-motion-slow}
\begin{aligned}
\dot {\bar{g}}_\mathrm{s}&=&f_\mathrm{s,g}^\mathrm{\odot}+f_\mathrm{s,g}^\mathrm{coorb}\,, \quad  \dot {\bar{G}}_\mathrm{f}&=&f_\mathrm{f,G}^\mathrm{coorb}\,;
\\
\dot {\bar{h}}_\mathrm{s}&=&f_\mathrm{s,h}^\mathrm{\odot}+f_\mathrm{s,h}^\mathrm{coorb}\,, \quad \dot {\bar{H}}_\mathrm{f}&=&f_\mathrm{f,H}^\mathrm{coorb}\,.
\end{aligned}
\end{equation}
Here $f_\mathrm{s,g}^\mathrm{\odot}$ is the rate of solar-driven variation in $\bar{g}_\mathrm{s}$ and is derived directly from partial differentiation of the Kozai Hamiltonian \eqref{slow-kozai}. The derivation of the coorbital potential contribution $f_\mathrm{f,G}^\mathrm{coorb}$ and $f_\mathrm{s,g}^\mathrm{coorb}$ is not so straightforward. Using the potential \eqref{slow-coorbital} and the equations of motion \eqref{co-euqation-motion}, we can express the variational rates of the relative variables $h_\mathrm{r}$ and $k_\mathrm{r}$ exerted by Himalia. The rates of the orbital elements of the particle are related to these rates through Eq. \eqref{both-equation-motion}. Then the transformation \eqref{trans2slow} allows us to derive the rates $f_\mathrm{f,G}^\mathrm{coorb}$ and $f_\mathrm{s,g}^\mathrm{coorb}$ in Eq. \eqref{equation-motion-slow}. Analogous relations exist for $\bar h_\mathrm{f}$ and $\bar H_\mathrm{s}$. In this way, we arrive at the above equations of motion \eqref{equation-motion-slow}. In essence, what we have is a four-dimensional system of $({\bar{g}}_\mathrm{s};{\bar{G}}_\mathrm{f}; {\bar{h}}_\mathrm{s};{\bar{H}}_\mathrm{f})$.

When describing the dynamics of the system, we still use the original orbital elements $\Delta\Omega=\Omega_\mathrm{P}-\Omega_\mathrm{H}$, $\Delta\varpi=\varpi_\mathrm{P}-\varpi_\mathrm{H}$, $e_\mathrm{P}$ and $i_\mathrm{P}$.


\subsection{General dynamics}
With the equations of motion above, we are able to integrate the system \eqref{equation-motion-slow} numerically to investigate its evolution\footnote{All integrations of the equations of motion derived from Kozai and/or coorbital theory are carried out with a 7th order Runge-Kutta method; the step-size is adjusted with an 8th order error estimate. The single step error tolerance is $10^{-14}$.}. We run test integrations and find that, at most locations in phase space, the angles $\Delta \Omega$ and $\Delta \varpi$ circulate with small amplitude oscillations in $e_\mathrm{P}$ and $i_\mathrm{P}$. This is not unexpected and results from the fact that, for positions in the $(a,e,i)$ space far from the SEPR, solar perturbations cause $\Delta \Omega$ and $\Delta \varpi$ to circulate fast, while the coorbital potential contributes minor changes to the precession rates and introduces small amplitude variations in $e_\mathrm{P}$ and $i_\mathrm{P}$.

However, when near the SEPR, Himalia can be as efficient as the Sun in changing the orbital evolution; we show this in the following subsections.


\subsection{Nodal resonance}
\label{sec-nodal}

\begin{figure}[h]
\begin{center}
	\includegraphics[width=0.8\textwidth]{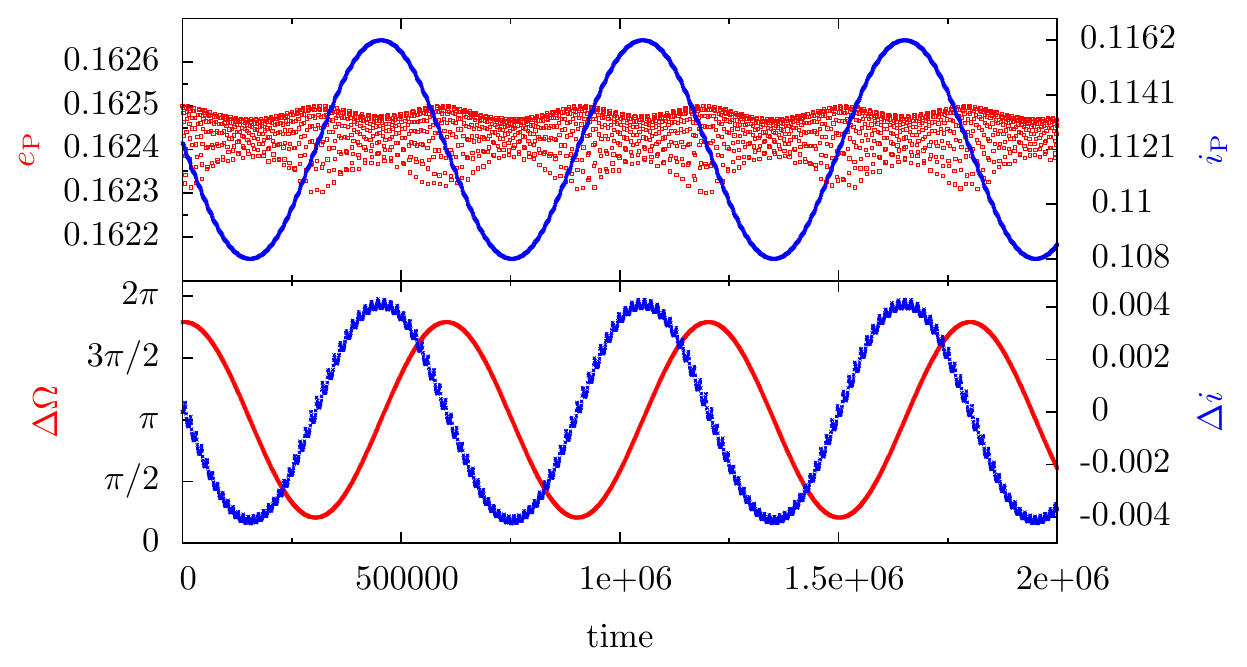}
\caption{An example of nodal libration from the four-dimensional system \eqref{equation-motion-slow}. Top panel: $e$ (red) and $i$ (blue); bottom panel: $\Delta \Omega$ (red) and the signed ``distance'' to the SEPR (blue); $x$-axis: time in our adopted unit. Since what matters is not the absolute ``distance'' but the relative position of the librating particle with respect to the SEPR, we use the quantity $\Delta i$ to represent this relation. $\Delta i$ is the distance from the particle to the point in the SEPR with the same $a$ and $e$. Thus $\Delta i>0$ means the test particle is above the SEPR and vice versa}
\label{hh-example-semi}
\end{center}
\end{figure}

In Fig. \ref{hh-example-semi}, we show an example of nodal libration in $\Delta \Omega$ in our integrations of Eq. \eqref{equation-motion-slow}. As observed in \citet{Christou2005}, the angle $\Delta \Omega$ librates around $\pi$. While the inclination oscillates in concert with the libration in the nodal difference, the eccentricity undergoes only small amplitude and short time-scale variations. In fact, the eccentricity is a quasi-conserved quantity here and in the two-dimensional system in Subsect. \ref{sec-phase-space}, it is actually constant. The timescale of nodal libration is of order $10^6$ ($\sim$ 1 Myr) and the changes in inclination are of order $10^{-3}$. To prove the postulate that the librator oscillates around the SEPR when in resonance, we plot the signed distance from the particle to the surface in the same figure. From the plot, it is clear that, when above the SEPR, $\Delta \Omega$ increases and vice versa\footnote{In fact, the node precesses with negative rates. Thus when above the SEPR, the test particle has smaller absolute precession rate.}. Inherently, when in nodal libration, the particle is restricted to be near the surface and to cross it regularly. This also shows the dominance of the Sun.

\subsection{Three additional slow angles}
\label{sec-two-resonance}

\begin{figure}[h]
\begin{center}
	\includegraphics[width=0.8\textwidth]{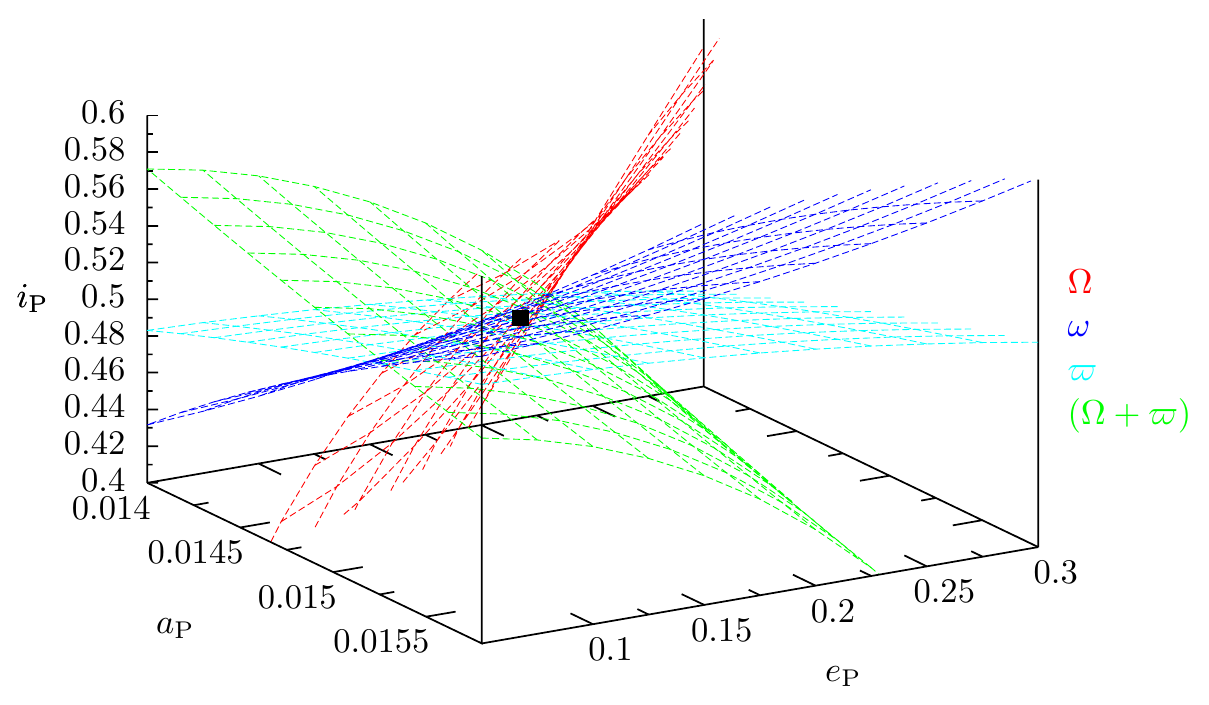}
\caption{The surfaces of equal precession rate in $ \Omega$, $ \omega= (\varpi-\Omega)$, $ \varpi$ and $ (\Omega+\varpi)$ in $(a,e,i)$ space. They are evaluated in the same way as Fig. \ref{equal-precession-hh}; the layout of the three axes and the viewing angle are also the same. Particles on these surfaces share the same precession rates in the angles $ \Omega$ (red), $ \omega= (\varpi-\Omega)$ (blue), $ \varpi$ (cyan) and $ (\Omega+\varpi)$ (green) with Himalia, respectively. The black square marks the position of Himalia. We deliberately choose this viewing angle so that, the normals to the surfaces appear perpendicular to the line of sight}
\label{equal-precession-all}
\end{center}
\end{figure}

Using the same method in Subsect. \ref{sec-analytical}, we find that the SEPR in terms of the pericentre $\dot {\varpi} (a,e,i)=\dot {\varpi}|_\mathrm{H}$ exists. Furthermore, the SEPRs corresponding to the angle combinations ${\mathrm d (\varpi-\Omega) \over \mathrm d t} (a,e,i)={\mathrm d \omega \over \mathrm d t} (a,e,i)={\mathrm d {\omega} \over \mathrm d t}|_\mathrm{H}$ and ${\mathrm d ({\Omega+\varpi}) \over \mathrm d t} (a,e,i)={\mathrm d ({\Omega+\varpi}) \over \mathrm d t}|_\mathrm{H}$ can be readily computed. We present them together with the SEPR in $\Omega$ in Fig. \ref{equal-precession-all}. The four surfaces share the same intersection curve.

\begin{figure}[h]
\begin{center}
	\includegraphics[width=0.8\textwidth]{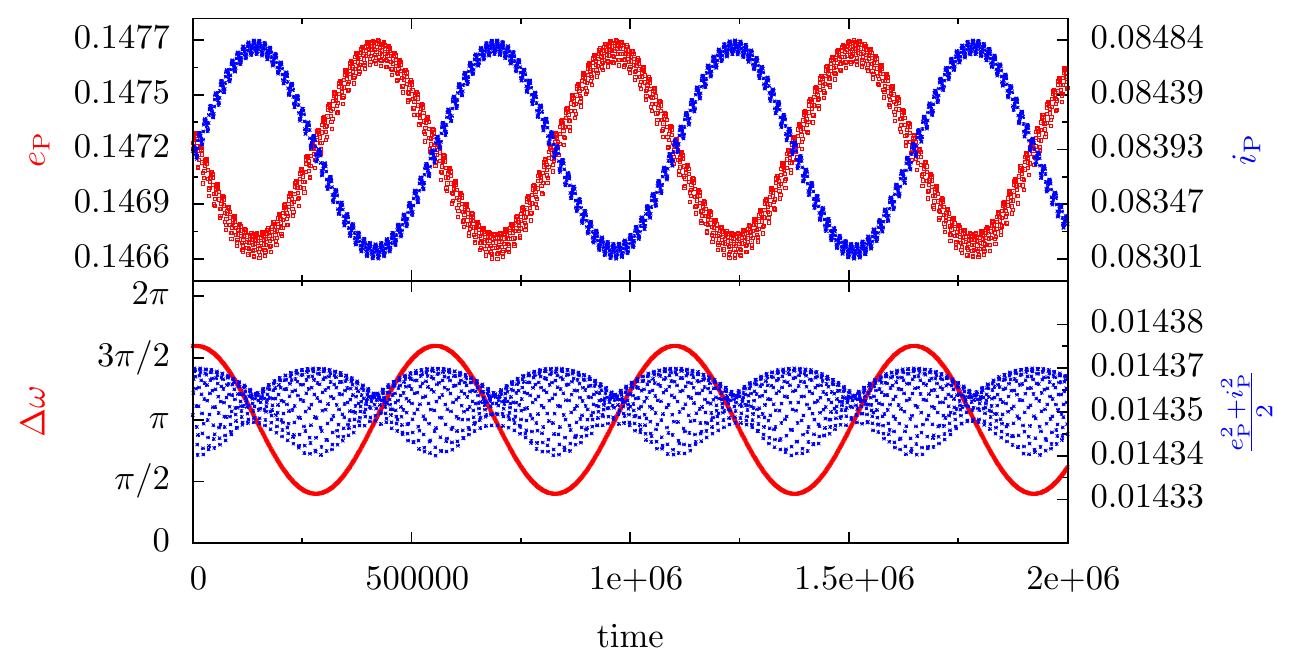}
\caption{An illustration of the resonance in the angle $\Delta \omega= \Delta (\varpi-\Omega)$. Top panel: $e$ (red) and $i$ (blue); bottom panel: $\Delta \omega$ (red) and the quasi-conserved quantity $(e_\mathrm{P}^2+i_\mathrm{P}^2)/2$ (blue); $x$-axis: time. This type of resonance is similar to the Kozai resonance}
\label{wmh-example-semi}
\end{center}
\end{figure}

\begin{figure}[h]
\begin{center}
	\includegraphics[width=0.8\textwidth]{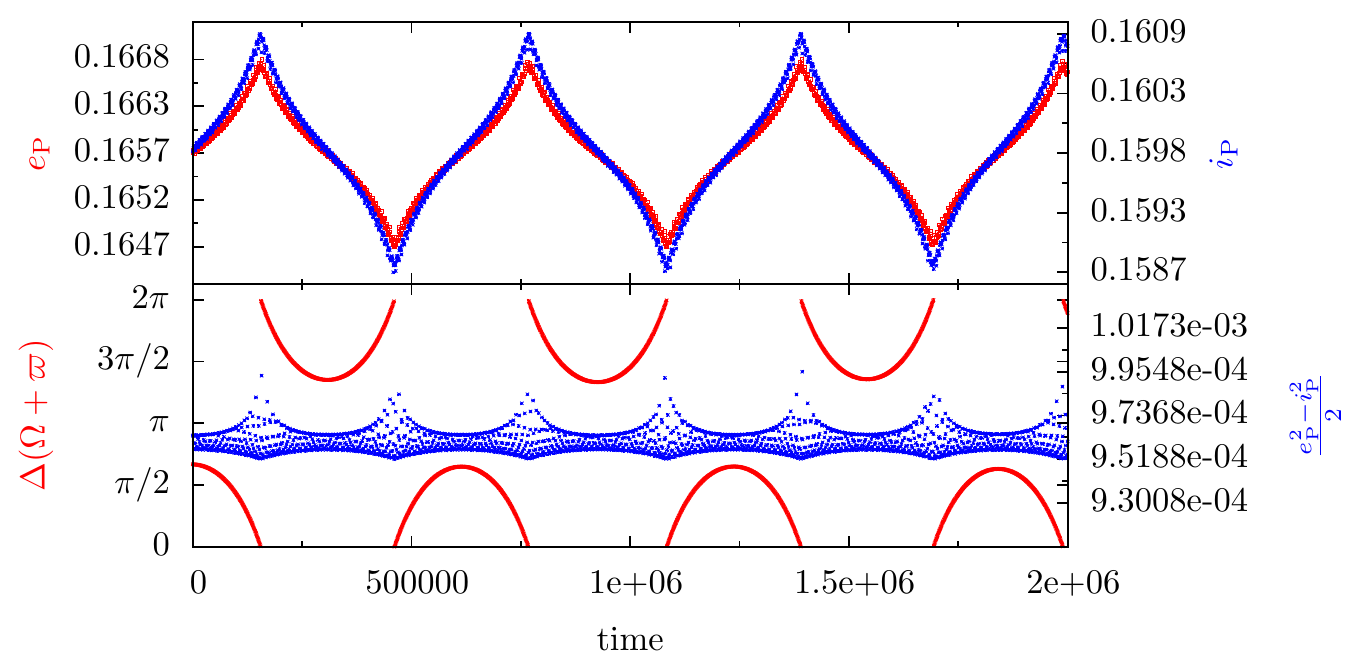}
\caption{A representative of the libration in $\Delta (\Omega+\varpi)$. Top panel: $e$ (red) and $i$ (blue); bottom panel: $\Delta (\Omega+\varpi)$ (red) and the quasi-conserved quantity $(e_\mathrm{P}^2-i_\mathrm{P}^2)/2$ (blue); $x$-axis: time}
\label{hpw-example-semi}
\end{center}
\end{figure}

\begin{figure}[h]
\begin{center}
	\includegraphics[width=0.8\textwidth]{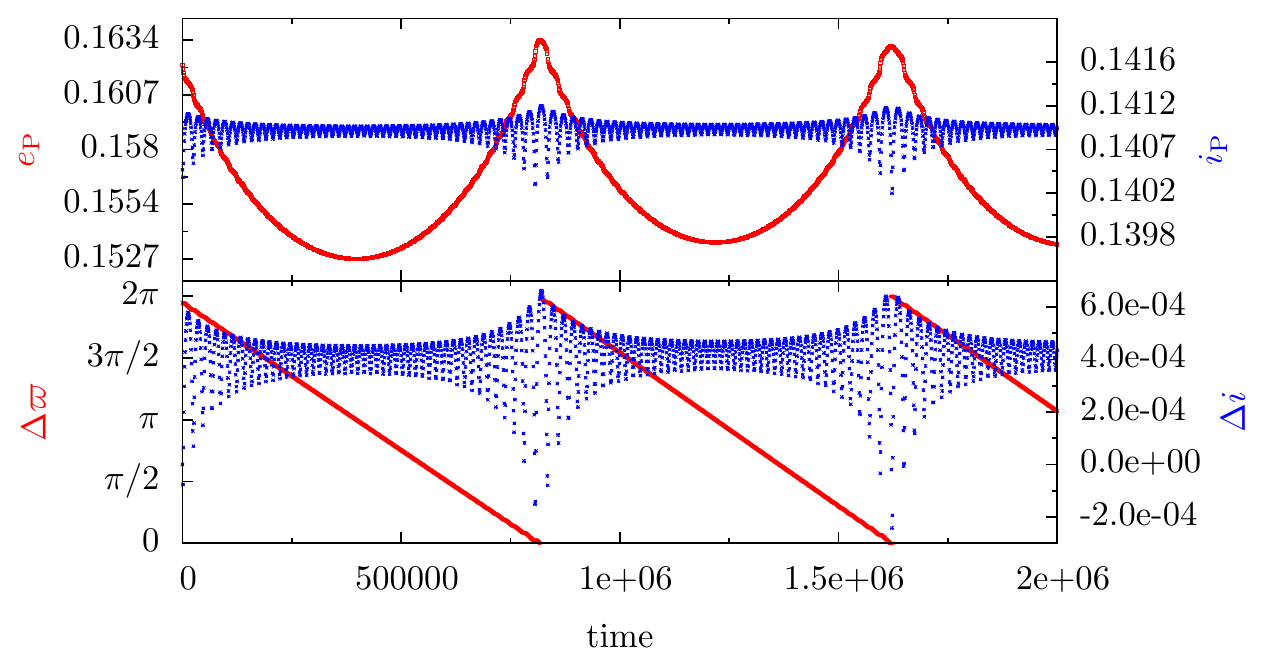}
\caption{Typical evolution of the slow circulation of the angle $\Delta \varpi$. Top panel: $e$ (red) and $i$ (blue); bottom panel: $\Delta \varpi$ (red) and the signed ``distance'' to the SEPR in $\varpi$ (blue); $x$-axis: time}
\label{ww-example-semi}
\end{center}
\end{figure}

Using our semianalytical model, we find that two of these three angles,  $\Delta \omega= \Delta (\varpi-\Omega)=\omega_\mathrm{P}-\omega_\mathrm{H}$ and $\Delta (\Omega+\varpi)=(\Omega_\mathrm{P}+\varpi_\mathrm{P})-(\Omega_\mathrm{H}+\varpi_\mathrm{H})$, can librate. We show examples of these resonances in Figs. \ref{wmh-example-semi} and \ref{hpw-example-semi}. Interestingly, we do not find libration in the angle $\Delta \varpi=\varpi_\mathrm{P}-\varpi_\mathrm{H}$; instead there is only slow circulation near its SEPR; see Fig. \ref{ww-example-semi}.

In the case of the $\Delta \omega$ resonance in Fig. \ref{wmh-example-semi}, while the angle librates about $\pi$, the variations of eccentricity and inclination are anti-correlated. This behaviour is reminiscent of the Kozai resonance \citep{Kozai1962}. As we will show in the next subsection, the quasi-conserved quantity is $(e_\mathrm{P}^2+i_\mathrm{P}^2)/2$, the component of the angular momentum along the $z$-axis; we plot this quantity in the same figure.

For the resonance of $\Delta (\Omega+\varpi)$ in Fig. \ref{hpw-example-semi}, the angle librates about $0$ while the eccentricity and inclination are correlated. The quasi-conserved quantity in this case is $(e_\mathrm{P}^2-i_\mathrm{P}^2)/2$.

What we have shown here are the examples where the angles librate. When outside libration but still close to the SEPRs, the angles circulate slowly and monotonically increase or decrease (since no crossing of the SEPR takes place now) with the actions evolving on the same timescale and in concert, like the case of the angle $\Delta \varpi$ described below.

For particles near the SEPR in $\varpi$, we detect no libration\footnote{ We generate 500 randomly distributed particles on the SEPR in $\varpi$ and integrate them for $2\times10^6$. Visual check shows no apsidal resonance.} thus we show an example of slow circulation in Fig. \ref{ww-example-semi}. When $\Delta \varpi$ circulates slowly, the eccentricity varies on the same timescale while the inclination experiences only short-period oscillations. In essence, inclination is the quasi-conserved quantity here. In this situation, the test particle cannot cross the SEPR in $ \varpi$, as is shown in the same figure. However, though the distance from the particle to the SEPR remains positive for most of the time, it does become negative for some time. According to the previous reasoning, the distance should not change its sign when in circulation. We suggest that, because of the coorbital potential, the SEPR from Kozai theory does not coincide with the SEPR of the full problem. As will be shown, the coorbital potential causes a small displacement in the surface. Thus although the distance in terms of the SEPR in Kozai potential changes the sign, the distance with respect to the SEPR controlled by both Kozai and coorbital potential does not. See Fig. \ref{equilibrium-nodal} for the influence of coorbital potential on the equilibrium points (i.e., the real SEPR). In Sect. \ref{sec-discussion} we discuss the reason why no examples of libration in $\Delta\varpi$ are observed here.

\subsection{Phase space properties for the three types of resonance}
\label{sec-phase-space}

\begin{figure}[h]
\begin{center}
	\includegraphics[width=0.8\textwidth]{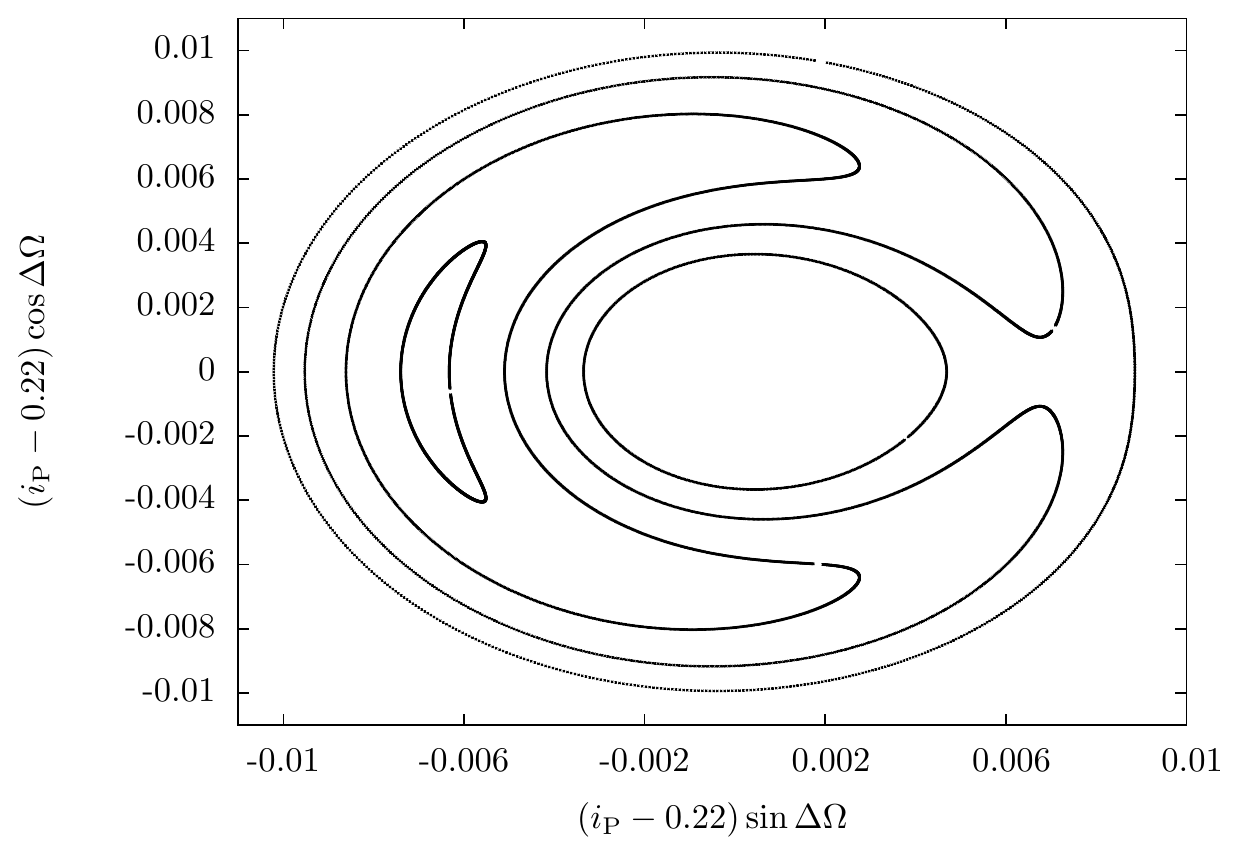}
\caption{Phase diagram for the nodal resonance in $\Delta \Omega$. All the orbits in this diagram share the same semimajor axis and the same conserved eccentricity, which are those of Lysithea. We subtract 0.22 from $i_\mathrm{P}$ mainly to highlight  the topological structure; the value of 0.22 has been empirically chosen. Cf. Fig. \ref{hh-example-semi}}
\label{phase-nodal}
\end{center}
\end{figure}

\begin{figure}[h]
\begin{center}
	\includegraphics[width=0.8\textwidth]{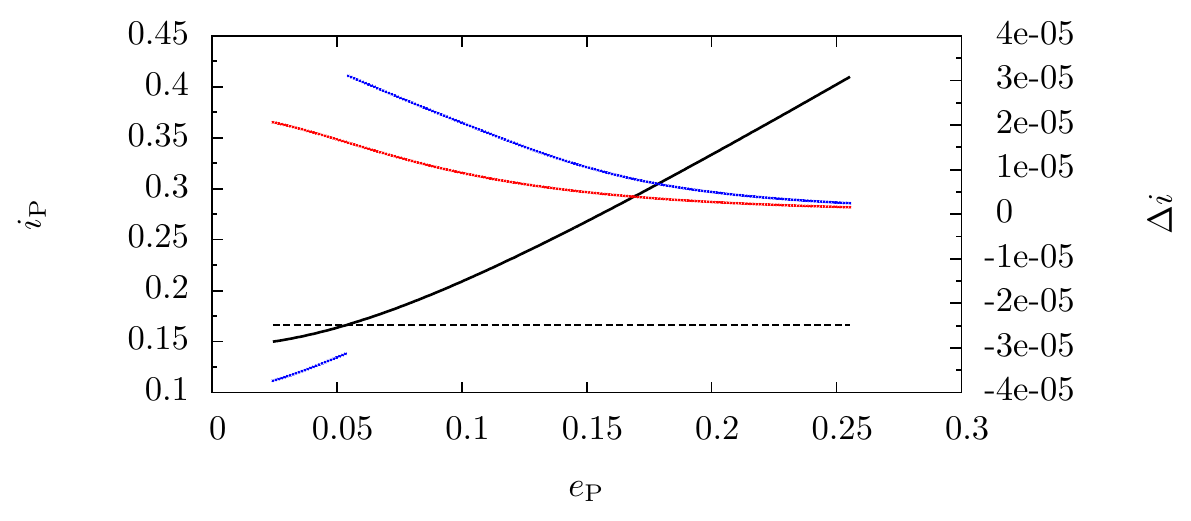}
\caption{The relative positions of the two equilibrium points of the nodal resonance. Left $y$-axis: the intersection of the SEPR in $\Omega$ with the $a=a_\mathrm{P}$ plane (black solid curve) and the inclination of Himalia (horizontal dashed line); right $y$-axis: distance from the equilibrium point with $\Delta \Omega=\pi$ (the stable one) to the black curve in terms of inclination (red) and distance from the equilibrium point with $\Delta \Omega=0$ (the unstable one, blue); $x$-axis: the varying eccentricity as a parameter}
\label{equilibrium-nodal}
\end{center}
\end{figure}

Having established that the three types of resonance exist in the four-dimensional system \eqref{equation-motion-slow}, we can reduce the system to two dimensions specifically to study the structure of individual resonances when the frequencies are well separated. We elaborate our approach for the nodal resonance and omit the details for the other two cases. Near the SEPR in $ \Omega$ and far from the common intersection curve of the four SEPRs, we have $\mathrm{d} \Delta \Omega / \mathrm{d}t \ll \mathrm{d} \Delta \varpi / \mathrm{d}t $. At these positions, the effect of the Sun in altering $\Delta \varpi$ far outweighs that of the coorbital interaction. Thus the evolution in $\Delta \varpi , e _\mathrm{P}$ is dominated by the Sun. In addition, under solar forcing \eqref{slow-kozai}, there is only (rapid) precession in $\Delta \varpi$ and no evolution in $e$. Hence, $\Delta \varpi$ is now a ``fast'' angle (compared to $\Delta \Omega$) and $e_\mathrm{P}$ is constant. To isolate the evolution in $(\Delta \Omega, i_\mathrm{P})$, we can perform the following averaging
\begin{equation}
\label{nodal-two-dimensional}
<\dot {\bar{h}}_\mathrm{s}>=f_\mathrm{s,h}^\mathrm{\odot}+{1\over 2\pi}\int_0^{2\pi} f_\mathrm{s,h}^\mathrm{coorb}\,\mathrm{d}\bar{g}_\mathrm{s}\,, \quad <\dot {\bar{H}}_\mathrm{f}>={1\over 2\pi}\int_0^{2\pi} f_\mathrm{f,H}^\mathrm{coorb}\,\mathrm{d}\bar{g}_\mathrm{s}
\end{equation}
to arrive at a two-dimensional system of $({\bar{h}}_\mathrm{s};{\bar{H}}_\mathrm{f})$ (Cf. Eq. \eqref{equation-motion-slow}). Here, all effects operating on timescales shorter than that of nodal libration are removed. We plot the evolution in $(\Delta \Omega , i _\mathrm{P})$ space in Fig. \ref{phase-nodal}; we elect to show the inclination relative to an empirically-determined value of 0.22 to highlight the amplitude of the variation. The topological structure of the resonance is clear: the system has two equilibrium points -- a stable one with $\Delta\Omega=\pi$ and an unstable one with $\Delta\Omega=0$; libration around the stable equilibrium point is allowed (which is the resonance we see). Hence, we suggest that with the coorbital potential, each point in the SEPR in Kozai potential is split into the two equilibrium points. This can be shown by setting the variation rates in \eqref{nodal-two-dimensional} to be zero. In Fig. \ref{equilibrium-nodal}, we show the positions of the two stationary points as functions of varying $e_\mathrm{P}$. The shift of the equilibrium points from the SEPR indicates the influence of the coorbital potential and is of order $10^{-5}$, reflecting the weakness of the coorbital interaction compared to the Kozai dynamics.

\begin{figure}[h]
\begin{center}
	\includegraphics[width=0.8\textwidth]{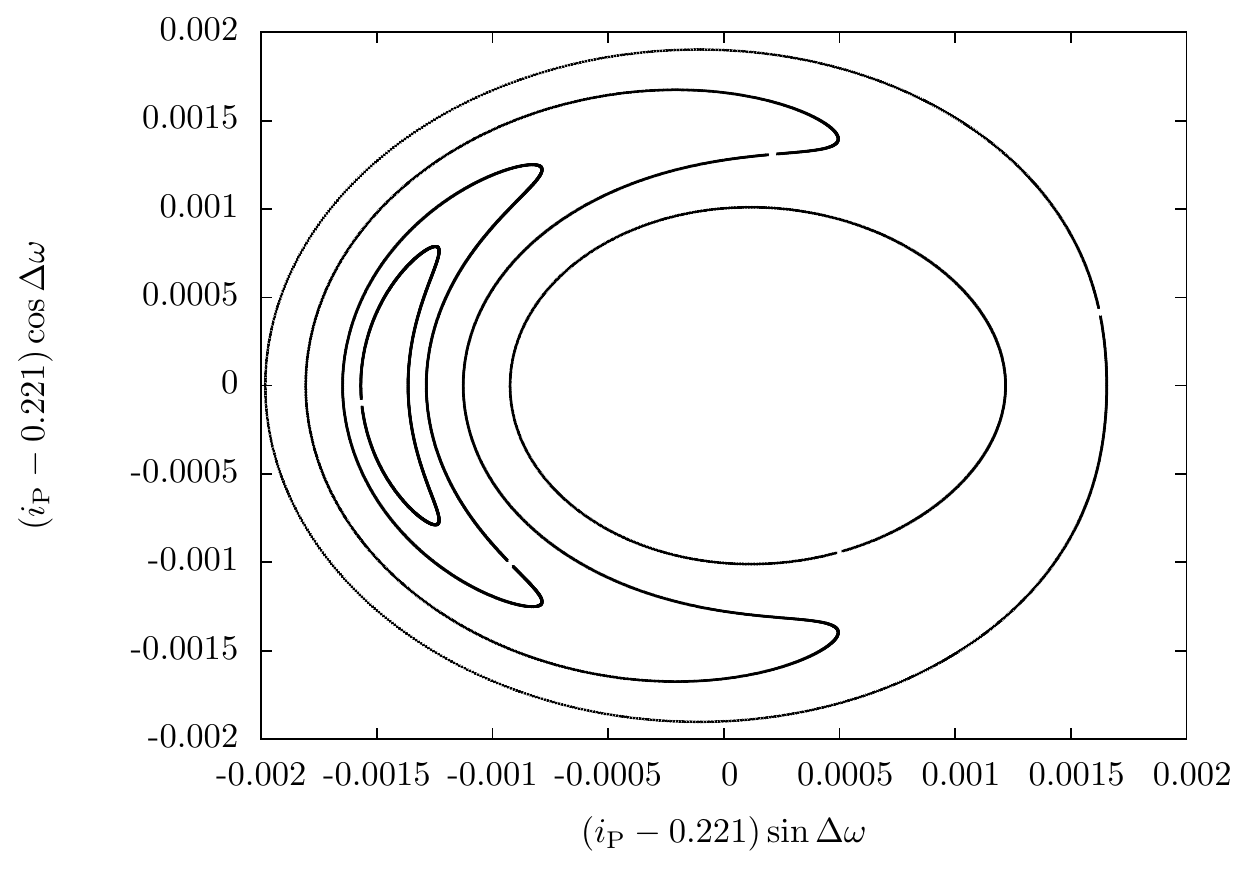}
\caption{Phase diagram for the resonance in $\Delta \omega=\Delta(\varpi-\Omega)$. Trajectories in this figure have the same semimajor axes and the same conserved quantity $H_{\Delta \omega}={ (e_\mathrm{P}^2+i_\mathrm{P}^2)/ 2}$. Intuitively, for any give $H_{\Delta \omega}$, we can solve for $e_\mathrm{P}$ and $i_\mathrm{P}$ so that, ${\mathrm{d} \Delta \omega /\mathrm{d}t}=0$ and this is a point on the corresponding SEPR in $ \omega$. However, the solution sometimes does not exist, which is the exact case for the $H_{\Delta \omega}$ prescribed by the elements of Lysithea. Thus we choose the point in the SEPR with the same semimajor axis and eccentricity as Lysithea. Then we use the $H_{\Delta \omega}$ of this point for all the trajectories plotted here. Cf. Fig. \ref{wmh-example-semi}}
\label{phase-wmh}
\end{center}
\end{figure}

\begin{figure}[h]
\begin{center}
	\includegraphics[width=0.8\textwidth]{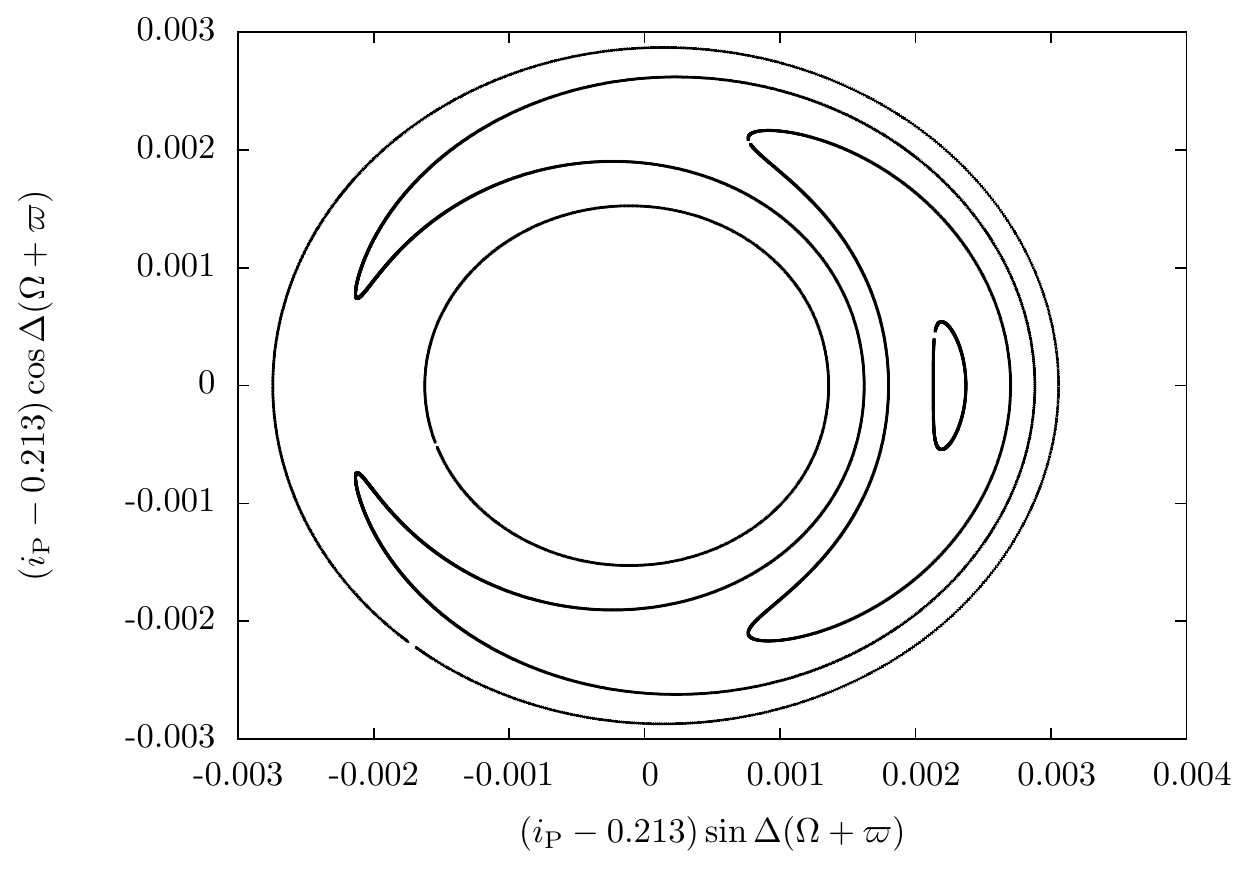}
\caption{Phase diagram for the resonance in $\Delta (\Omega+\varpi)$. Again, the curves possess the same semimajor axes and the same conserved quantity $H_{\Delta (\Omega+\varpi)}={ (e_\mathrm{P}^2-i_\mathrm{P}^2)/2}$ that is chosen in a similar procedure as in Fig. \ref{phase-wmh}. Cf. Fig. \ref{hpw-example-semi}}
\label{phase-hpw}
\end{center}
\end{figure}

Using similar techniques, we generate analogous plots for the resonances involving the angles $\Delta \omega=\Delta(\varpi-\Omega)$ and $\Delta (\Omega+\varpi)$ in Figs. \ref{phase-wmh} and \ref{phase-hpw}. In dealing with these two types of resonance, transformations to fully isolate the ``slow'' evolution are needed. The conserved quantities in the two-dimensional systems that are the angular momenta conjugate to the ``fast'' angles in the four-dimensional systems are
\begin{equation}
H_{\Delta \omega}={ e_\mathrm{P}^2+i_\mathrm{P}^2\over 2}
\end{equation}
for the resonance in $\Delta \omega$ and
\begin{equation}
H_{\Delta (\Omega+\varpi)}={ e_\mathrm{P}^2-i_\mathrm{P}^2\over 2}
\end{equation}
for the resonance in $\Delta (\Omega+\varpi)$, respectively.

Similar to the concept resonance width, we use the amplitude of the oscillation of inclination near the separatrix between libration and circulation as an indicator of the strength of the resonance. It is then observed that, the resonance in $\Delta \Omega$ is the strongest, which may be the reason why it was the first to be identified \citep{Christou2005}. In Fig. \ref{phase-nodal}, we observe that the amplitude of the variation of inclination in $\Delta \Omega$ resonance can be as high as $\approx 0.005$. The strengths of the resonances in $\Delta \omega$ and $\Delta (\Omega+\varpi)$ are similar to each other: the amplitudes of the oscillation of inclination are $\lesssim 0.001$.

Since there is apparently no libration in $\Delta \varpi$ in the semianalytical model, the structure of the phase space for the circulation should be trivial. Hence, we omit its phase diagram. Similar with the reasoning above, in the case where the slow angle is $\Delta \varpi$, the conserved quantity is $i_\mathrm{P}$.


\section{$N$-body simulations}
\label{sec-n-body}

Mainly due to the expansion in inclination, the semianalytical approach above is formally not accurate for the real Himalia group although it does serve to illustrate the dynamics. We also need to confirm whether these resonances identified by our semianalytical model survive in the full problem. For these reasons we perform extensive $N$-body simulations\footnote{All $N$-body simulations are implemented with the general Bulirsch-Stoer algorithm in the MERCURY package \citep{Chambers1999,Hahn2005} with a tolerance $10^{-12}$. Over an integration time of $10^8$ yr, the errors in energy and angular momentum are both of order $10^{-10}$.} to explore the parameter space of the real satellites.

The most important questions we wish to answer here are (i) how well the properties of these resonances are represented by the semianalytical model and, (ii) the possibility for a fictitious member of the family to be involved in such resonances. To explore these, an ensemble of massless test particles are integrated with Jupiter, the Sun and Himalia.

\citet{Beauge2007} used the metric
\begin{equation}
d^2=C_a (\Delta a /a)^2+ C_e (\Delta e)^2 + C_i (\Delta \sin i)^2
\end{equation}
to represent the distance from a satellite to the centre of a family in the proper element space ($C_a, C_e$ and $C_i$ are constant for a specific irregular satellite family). They proposed that for the Himalia family, a satellite with $ d \lesssim $ 320 m/s could be regarded as a member\footnote{Note that, when they published the paper, four satellites: Himalia, Elara, Leda and Lysithea were confirmed as group members. The recently recovered fifth member, Dia, has the largest semimajor axis and eccentricity. Our simple experiments show that the criterion has to be updated to $ d \lesssim $ 510 m/s to include Dia. In this paper, we stick to the results of \citet{Beauge2007}.}. If generating test particles with the above criterion only, we produce an ensemble of particles with the eccentricity in the range $(0.03,0.3)$ and inclination between $(23^\circ,33^\circ)$. These are too large, considering that the corresponding ranges for the real family are about $(0.11,0.22)$ and $(27^\circ,29^\circ)$. To represent better the real family and to enhance the number density of particles in $(a,e,i)$ space, we further require the eccentricity to be less than 0.1 and the inclination less than $2^\circ$ from the family mean values; these two values are approximately twice those of the actual family. We do not apply specific restrictions on the semimajor axes as long as they fulfil the distance criterion. The final distribution of the test particles is: $a\in (0.745 \mathrm{AU}, 0.795 \mathrm{AU})$, $e\in (0.06, 0.26)$ and $i \in (26 ^\circ, 30^\circ)$. We generate 1000 such particles. Strictly speaking, the metric is valid for proper elements. We have applied it to osculating elements for simplicity. The particles, along with Jupiter, Sun and Himalia\footnote{In the $N$-body simulations, the initial osculating elements of the real objects are taken from the JPL HORIZONS System \url{http://ssd.jpl.nasa.gov/?horizons}.}, are integrated for 100 Myr (about $5.3\times 10^7$ in our time unit). Our integrations take place in a jovi-centric reference system. Following \cite{Hinse2010}, we have changed the longitudes of pericentre by adding a $\pi$ so that, the elements are correct in the jovi-centric frame. The solar eccentricity $e_\odot$ has been set to zero.

Our $N$-body runs confirm the existence of the resonances in $\Delta \Omega$, $\Delta \omega$ and $\Delta (\Omega+\varpi)$ and discover another one involving the angle $\Delta \varpi$. We present examples of them in the following subsection. In Subsect. \ref{n-statistics-sec}, a statistical study of the resonances to address question (ii) stated in the beginning of this section is presented.


\subsection{Resonances in $N$-body simulations}
\label{sec-n-resonance}

\begin{figure}[h]
\begin{center}
	\includegraphics[width=0.8\textwidth]{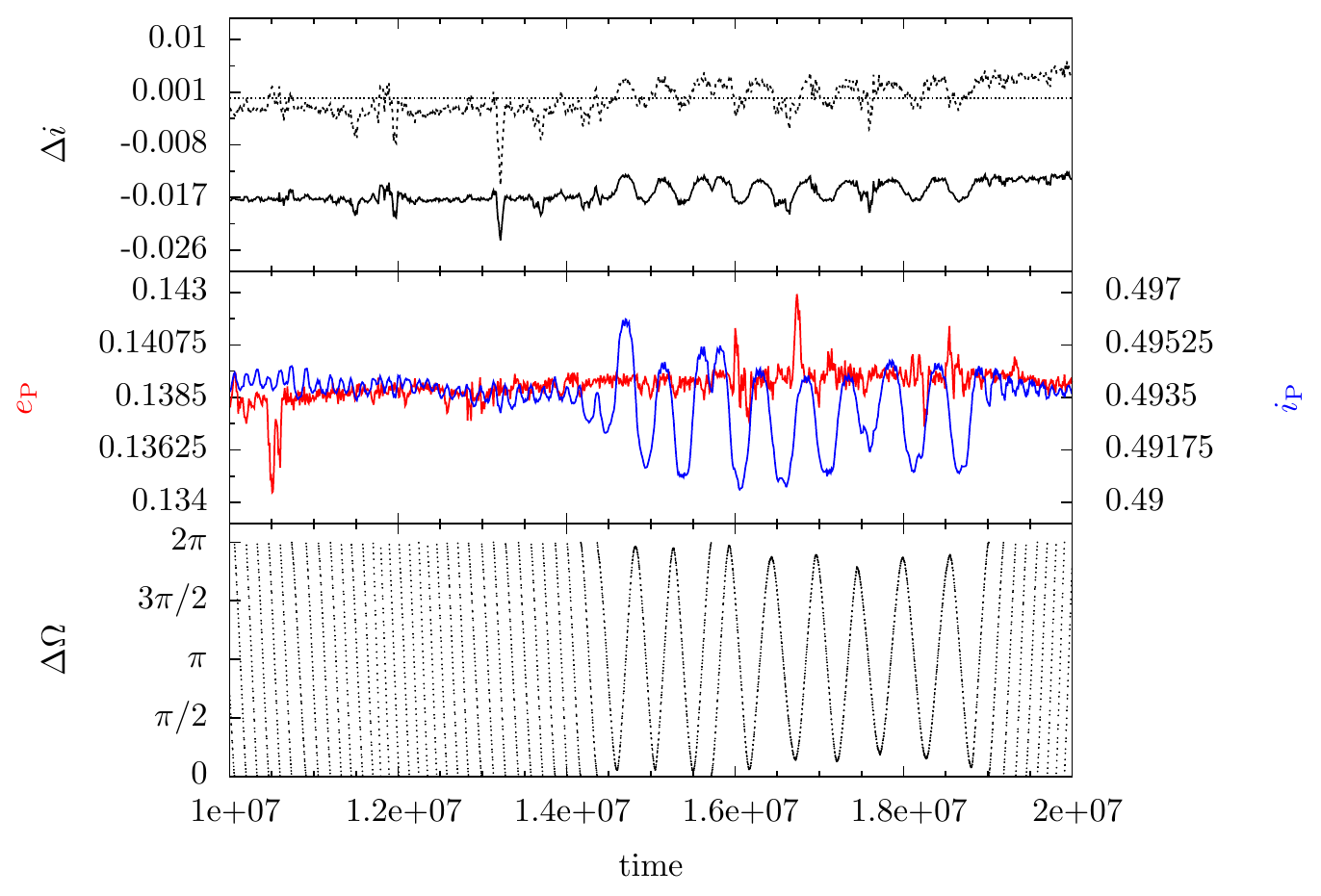}
\caption{An example of the libration in $\Delta \Omega=\Omega_\mathrm{P}-\Omega_\mathrm{H}$ from the $N$-body simulations. Cf. Figs. \ref{hh-example-semi} and \ref{phase-nodal}. Top panel: distance $\Delta i$ measured as $\Delta i$ to the SEPR; middle panel: $e$ (red) and $i$ (blue); bottom panel: $\Delta \Omega$; $x$-axis: time in our adopted unit. It is evident that, when in libration, $i$ (but not $e$) is modulated on a timescale of $10^6$. The distance $\Delta i$ has the same meaning as in Fig. \ref{hh-example-semi}: the solid curve is the distance to the SEPR from semianalytical model \eqref{constant-kozai}; the dashed one is related to the fitted surface (Cf. Eq. \eqref{fit-surface} and Fig. \ref{elements-resonance}); the horizontal dotted line marks a value of zero for $\Delta i$}
\label{n-hh}
\end{center}
\end{figure}

In Fig. \ref{n-hh}, we show a typical example of a test particle in (temporary) nodal libration with Himalia. As in \cite{Christou2005}, the quantities $e$, $i$ are averaged in a $\sim 5\times10^4$ time window to suppress the short-period oscllations. Some of these fluctuations come from the Kozai cycle.

The $N$-body simulations agree well with the semianalytical model in reproducing the general features of the nodal resonance, e.g., the centre and period of libration in $\Delta \Omega=\Omega_\mathrm{P}-\Omega_\mathrm{H}$ ($\pi$ and $10^6$ respectively), the corresponding oscillation in inclination and the failure of the resonance to excite the eccentricity. Inspecting the example in Fig. \ref{n-hh} in detail, it is found that, when the signed distance to the SEPR in $(a,e,i)$ space, measured in $\Delta i$ (top panel, the solid curve), is around the peak,  $\Delta \Omega$ increases and when $\Delta i$ is near the trough, $\Delta \Omega$ decreases. This is consistent with the semianalytical example (Fig. \ref{hh-example-semi}) in the sense that when the particle moves above the SEPR with higher inclination, it has higher nodal precession rate than Himalia (while its absolute value is lower since the rate is negative), and vice versa; as  $\Delta \Omega$ approaches zero, the particle is pushed to cross the surface, causing the libration. However, in this example, the distance does not oscillate exactly around zero, but slightly below it. We suggest, this is because of the simplifications we made in Sect. \ref{sec-analytical} in constructing the semianalytical model and especially, in calculating the SEPR; in the full problem, such oscillations should be about zero. Alternatively, if we use the fit to the data from the $N$-body simulations (see Subsect. \ref{n-statistics-sec} and particularly, Eq. \eqref{fit-surface} and Fig. \ref{elements-resonance}), the result turns out to be better. When in libration, the distance $\Delta i$ (the dashed curve in the top panel) does oscillate around zero. Also apparent is that, before about $1.4\times10^7$, the distance remains smaller than zero and $\Delta \Omega$ keeps decreasing; then from roughly $1.9\times10^7$, the angle, after passage through libration, becomes increasing, with the distance above zero.

\begin{figure}[h]
\begin{center}
	\includegraphics[width=0.8\textwidth]{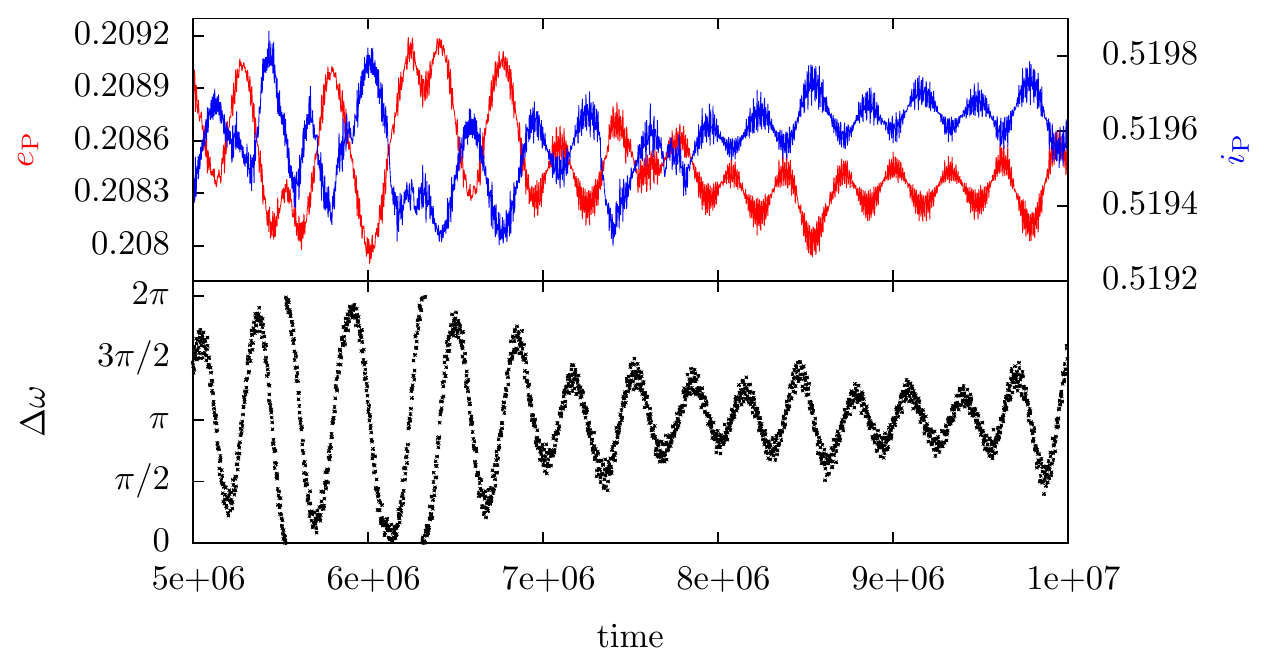}
\caption{An example of the libration in $\Delta \omega=\omega_\mathrm{P}-\omega_\mathrm{H}$. Cf. Figs. \ref{wmh-example-semi} and \ref{phase-wmh}. Top panel: $e$ (red) and $i$ (blue); bottom panel: $\Delta \omega$; $x$-axis: time}
\label{n-wmh}
\end{center}
\end{figure}

In Fig. \ref{n-wmh} we show an example of the resonance in $\Delta \omega=\Delta(\varpi-\Omega)=\omega_\mathrm{P}-\omega_\mathrm{H}$. As observed by our integrations of the semianalytical model (Fig. \ref{wmh-example-semi}), the angle $\Delta \omega$ librates around $\pi$. The variation of the eccentricity and inclination is anti-correlated.  This is reminiscent of the Kozai cycle.

\begin{figure}[h]
\begin{center}
	\includegraphics[width=0.8\textwidth]{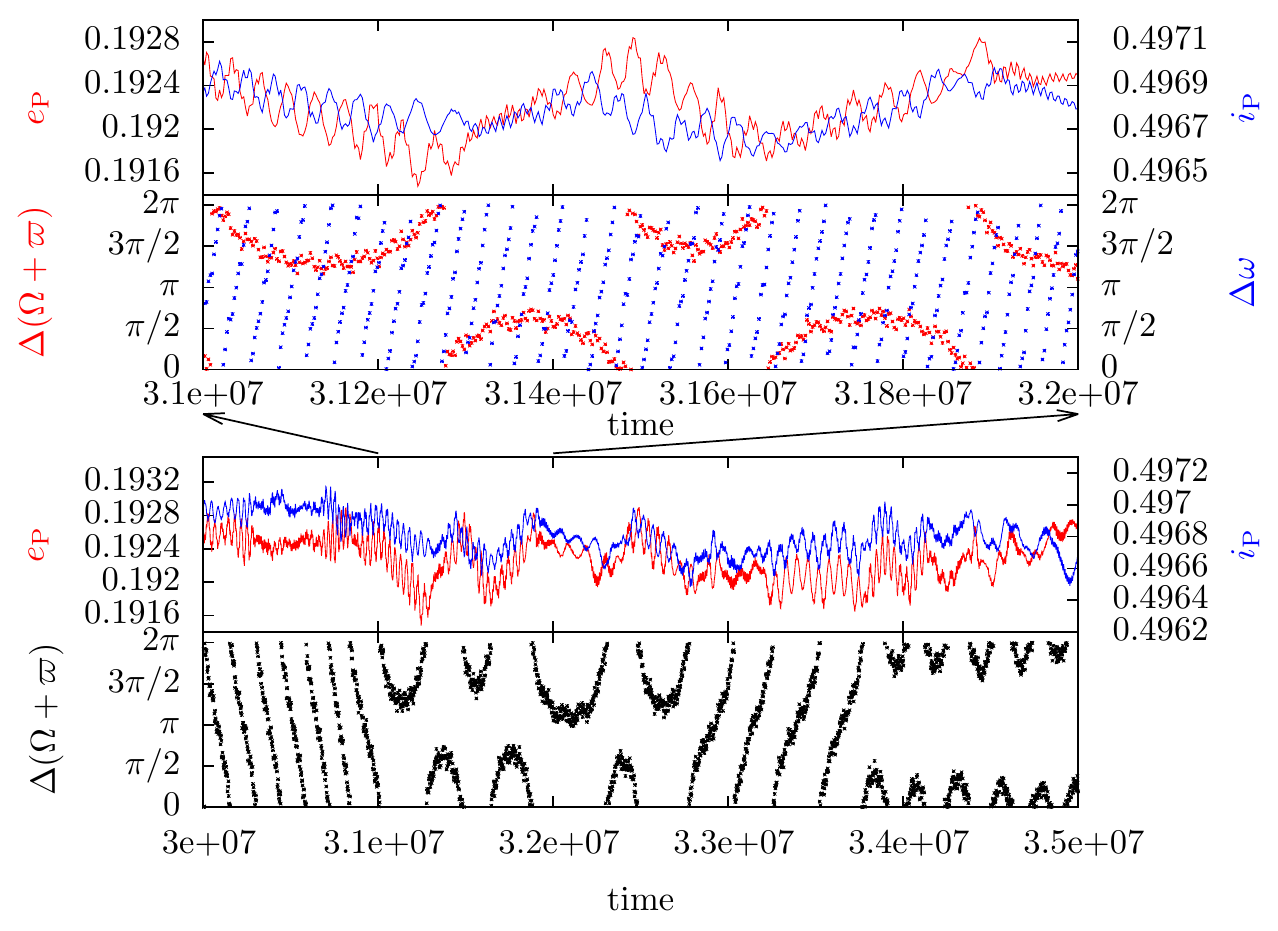}
\caption{An example of the libration in $\Delta (\Omega+\varpi)=(\Omega_\mathrm{P}+\varpi_\mathrm{P})-(\Omega_\mathrm{H}+\varpi_\mathrm{H})$ from the $N$-body simulations. Cf. Figs. \ref{hpw-example-semi} and \ref{phase-hpw}. In the lower two panels: top panel: $e$ (red) and $i$ (blue); bottom panel: $\Delta (\Omega+\varpi)$; $x$-axis: time. The two upper panels represent enlargements of each of the bottom two panels: top panel: $e$ (red) and $i$ (blue); bottom panel: $\Delta (\Omega+\varpi)$ (red) and $\Delta \omega$ (blue); $x$-axis: time}
\label{n-hpw}
\end{center}
\end{figure}

We show an example for the resonance in $\Delta (\Omega+\varpi)=(\Omega_\mathrm{P}+\varpi_\mathrm{P})-(\Omega_\mathrm{H}+\varpi_\mathrm{H})$ as the lower panels in Fig. \ref{n-hpw}. This test particle evolves from circulation into libration, leaves libration and enters libration again. Here, the evolution of eccentricity and inclination is correlated: maxima (or minima) of $e$ and $i$ occur at the same time. Interestingly, for this particular example, the rate of circulation of the angle $\Delta \omega$ is not much faster than that of $\Delta (\Omega+\varpi)$. In the upper two panels, we show an enlargement of each of the lower two panels of Fig. \ref{n-hpw}. The short-period and anti-correlated evolution of eccentricity and inclination appear superposed on the long-period correlated evolution. We believe that such short-period effects are related to the circulation of $\Delta\omega$. Since the resonances are so close together in $(a,e,i)$ space (see Figs. \ref{equal-precession-all} and \ref{elements-resonance}), it is not surprising such slow circulation of an angle exists inside the libration region of another angle.

\begin{figure}[h]
\begin{center}
	\includegraphics[width=0.8\textwidth]{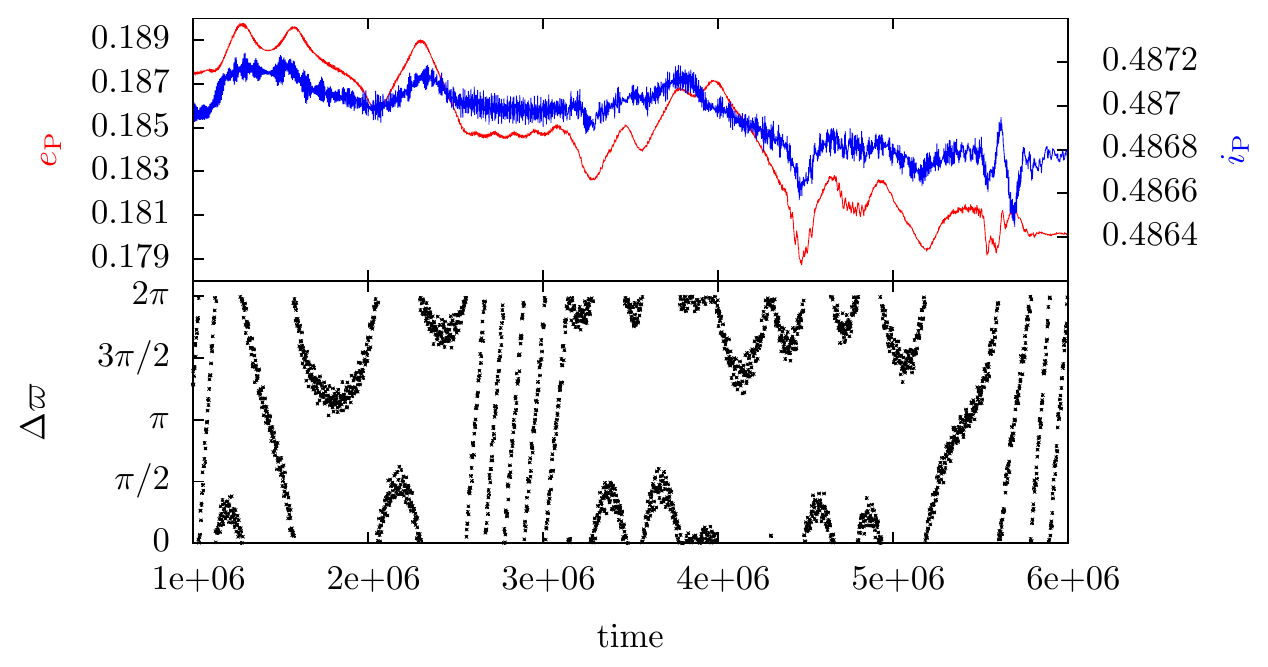}
\caption{An example of the libration in $\Delta \varpi=\varpi_\mathrm{P}-\varpi_\mathrm{H}$. Cf. Fig. \ref{ww-example-semi}. Top panel: $e$ (red) and $i$ (blue); bottom panel: $\Delta \varpi$; $x$-axis: time}
\label{n-ww}
\end{center}
\end{figure}

Apart from the three types of resonance that appear in the semianalytical model and confirmed by our N-body runs here, we observe another type involving the angle $\Delta \varpi=\varpi_\mathrm{P}-\varpi_\mathrm{H}$. We illustrate it with an example in Fig. \ref{n-ww}. The resonance in this angle seems to be the weakest as the resonant angle and action do not behave as regular as those in other three resonances. For the resonance in $\Delta \varpi$, we expect that, while the inclination remains nearly constant on the long term, the eccentricity oscillates periodically in accordance with the libration in $\Delta \varpi$. However, the resonance appears to be highly unstable in the sense that, (i) the libration in $\Delta \varpi$ is intermittent and that (ii) the oscillation of eccentricity does not correlate well with the variation of the angle. In Sect. \ref{sec-discussion}, we will further explain this. The libration in $\Delta \varpi$ will be referred to as ``apsidal libration'' or ``apsidal resonance'' thereafter.

\subsection{Statistics of the $N$-body simulations}
\label{n-statistics-sec}

Using the high number of libration episodes observed in our simulations, we can draw some statistical conclusions.

Out of the 1000 test particles, 37 are found to be in nodal resonance for longer than about $10^6$ (by visual inspection); among those 37, 32 show libration for time spans $\gtrsim 3 \times 10^6$. A duration of $10^6$ corresponds to approximately two cycles of the nodal libration.

To identify episodes of libration automatically in the time series, we implement two different approaches. The first approach uses the discrete Fast Fourier Transform (FFT) algorithm in the package ``Fastest Fourier Transform in the West'' \citep{Frigo2005} to extract the periods and amplitudes of the periodic terms in the time series for $\Delta \Omega$ and $i$; then we identify and compare the dominant periods of the two. In the Fourier spectrum, the distribution of power as a function of period around the peak satisfies the normal distribution
\begin{equation}
P(x)= A\, e^{ -\frac{(x-\mu_0)^2}{2\sigma_0 ^2} }\,,
\end{equation}
where $P(x)$ is the strength at period $x$, $A$ a scaling factor, $\mu_0$ the centre of the normal distribution and $\sigma_0$ its dispersion. Taking the logarithm of both sides, we have
\begin{equation}
\log {P(x)}=\log A -\frac{(x-\mu_0)^2}{2\sigma_0 ^2}.
\end{equation}
We fit this quadratic function to the data to extract the strongest period $\mu_0$ and its strength $P(\mu_0)$. We have tried Gaussian fitting of the defined original strength and it often does not converge. To improve the procedure, we instead weigh our data in an iterative process. The weight is designed so that, points with periods closer to $\mu_0$ and/or with larger strength $P$ are more important; the weight function is
\begin{equation}
\label{weight-function}
W_n={1+(P_n/\bar P)^M\over 1+ {|x_n - \mu_0 |/ \Delta x}}\,,
\end{equation}
in which $(x_n,P_n)$ is a (period, strength) combination from the FFT analysis, $W_n$ its weight, $\bar P$ the mean strength of all periods, $\Delta x$ the total coverage in period and $M$ a free parameter that controls the importance of the periods with different strength $P$. The value of $\mu_0$ is updated after each iteration. This process continues until either $\mu_0$ and $P(\mu_0)$ converge or do not. The analysis of inclination is complicated as it often shows multiple local extrema. Thus for inclination, if $M$ is too large, $\mu_0$ and $P(\mu_0)$ can converge towards a local extremum. For $\Delta \Omega$, the fluctuation is smaller, and a larger $M$ allows us to achieve a better estimate of the strongest period. From experiments, we choose $M=20$ for $\Delta \Omega$ and 4 for $i$; we use $(x_n,P_n)$ only when $x_n\in(3\times10^5,9\times10^5)$ to reduce the influences of irrelevant periods. In addition, we apply the criterion that if the fitted $P(\mu_0)$ is too small, we do not recognise the peak as real. In this way, we can identify all the 32 resonant particles with total time in libration $\ge3 \times 10^6$, without any false positives.

\begin{figure}[h]
\begin{center}
	\includegraphics[width=0.8\textwidth]{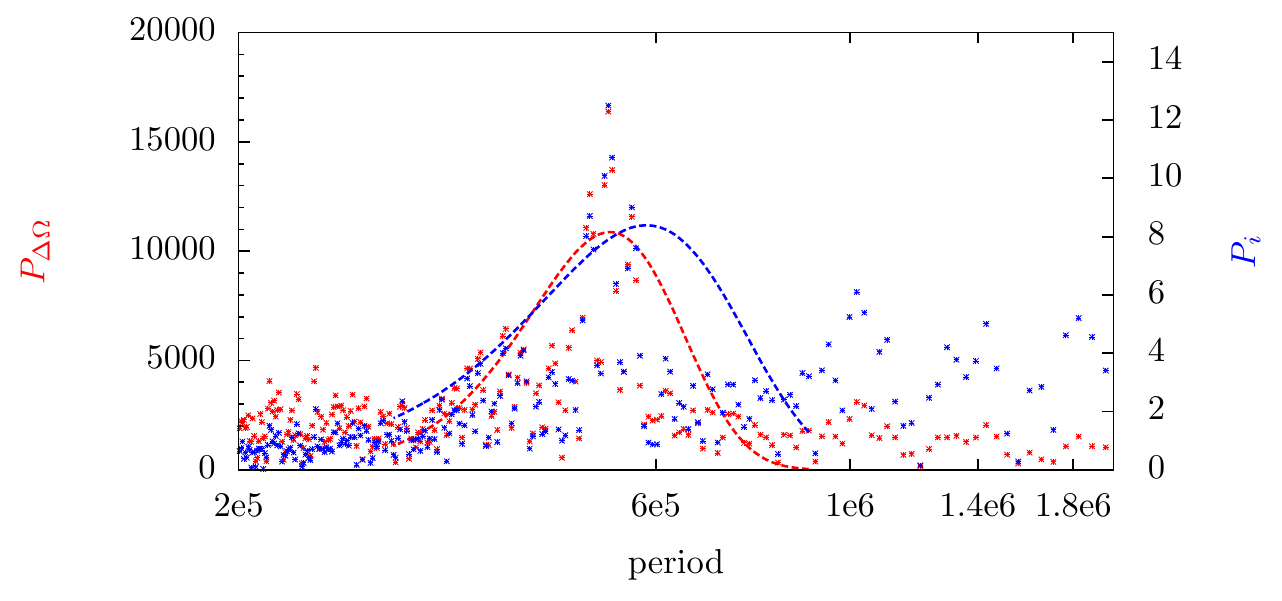}
\caption{The FFT analysis results and the represented fit in period-strength coordinates for $\Delta \Omega$ and $i$. Left $y$-axis: the FFT algorithm generated strength (point, red) and fitted strength (dashed curve, red) for $\Delta \Omega$; right $y$-axis: the FFT algorithm generated strength (point, blue) and fitted strength (dashed curve, blue) for $i$; $x$-axis: period. This is the analysis of the same particle as in Fig. \ref{n-hh}}
\label{fit-example}
\end{center}
\end{figure}

Shown in Fig. \ref{fit-example} is the FFT analysis and the fitting results of the same test particle as in Fig. \ref{n-hh}. Out of the entire integration time of $5.3\times10^7$, the total duration in libration is roughly $7\times10^6$. This is $(7\times10^6)/(5.3\times10^7)\approx 0.13$ of the entire integration and it generates a strong enough signature in FFT analysis that enables us to detect the libration. The period of nodal libration from the fit is $5.31\times 10^5$ for this particle whereas the average period of the 32 cases is $5.73\times 10^5$. As stated above, the determination of libration period in terms of $\Delta \Omega$ is more reliable owing to the higher value of $M$ in the weight function \eqref{weight-function}; the two periods are from the analysis of $\Delta \Omega$.

However, the method described above does not work well for the resonances in $\Delta \omega$, $\Delta (\Omega+\varpi)$ and $\Delta \varpi$. We suggest this issue mainly results from the different behaviour of the resonances. For instance, in Fig. \ref{n-hpw} for the resonance involving $\Delta (\Omega+\varpi)$, we see that the behaviour in the time range $(3.1\times 10^7,3.25\times10^7)$ and in $(3.36\times10^7,3.5\times10^7)$ are visually different. Indeed, in the former, the particle is closer to the separatrix with larger amplitude and longer period compared to the latter one where the particle is near the libration centre. In nodal libration, the librator tends to stay near the separatrix (see Fig. \ref{n-hh}); this may help to generate strong FFT signals\footnote{The particle in Fig. \ref{most-stable} described below is not a typical example: for a librator to be stable for $5.3\times10^7$, it has to lie well within the resonance. Most librators only pass through libration.}. Also these resonances are weaker with smaller amplitudes in the oscillations of $e$ and/or $i$ compared to the nodal resonance. These two factors may cause the failure of the above method to identify these episodes of libration for the three resonances.

To provide a universal approach valid for all the resonances, we introduce the following method which detects libration depending on the behaviour of the angles only.

The basic idea is that principally, when an angle is in libration, it cannot cross the point that is $\pi$ away from the libration centre and it must cross the libration centre regularly. For instance, in the case of nodal libration (i.e., when the libration centre is $\pi$), the angle $\Delta \Omega$ should not cross 0 and it is supposed to cross $\pi$ periodically. If this state of affairs persists, we recognise it as a libration event. Additionally, we require the frequency of crossing the libration centre and the rate of the librating angle to be within plausible ranges. Therefore, the addition of these filters reduces significantly the number of false positives. In testing the method for nodal libration, we empirically choose the parameters so that it retrieves all the 32 librators found by the FFT-based method and without false positives. In this respect, the two methods are mutually consistent.

With this method, we find 32, 29, 19 and 11 particles in resonances involving $\Delta \Omega$, $\Delta \omega$, $\Delta (\Omega+\varpi)$ and $\Delta \varpi$, respectively.

\begin{figure}[h]
\begin{center}
	\includegraphics[width=0.8\textwidth]{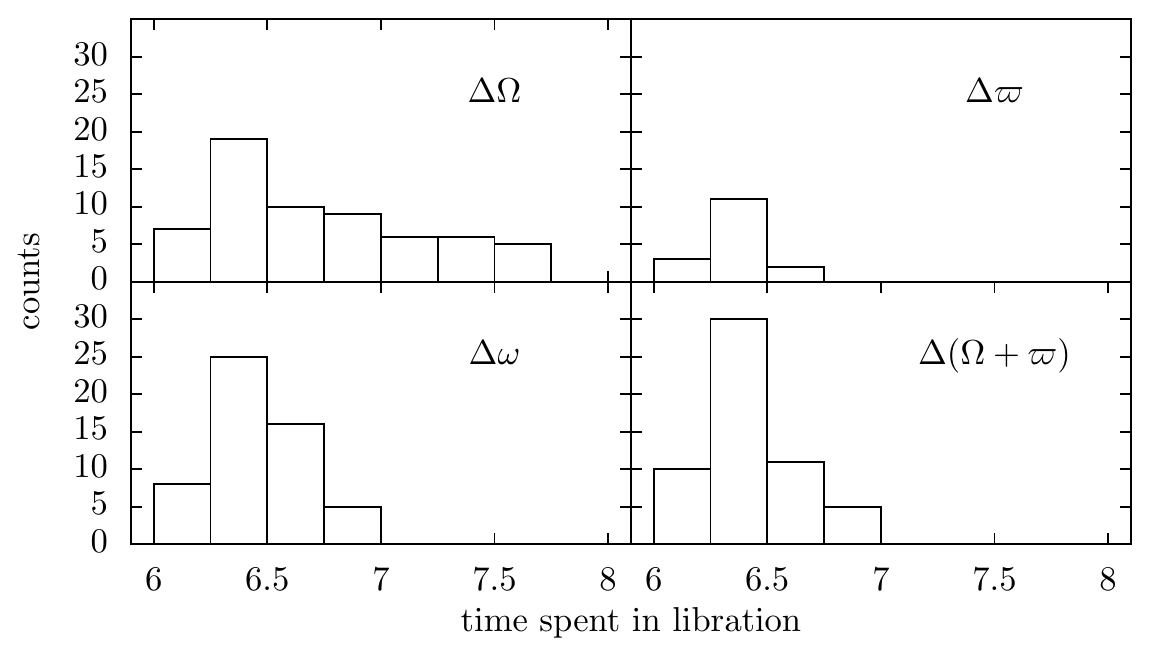}
\caption{The distribution of time spent in resonance of each libration event. Top-left: resonance involving $\Delta \Omega$; bottom-left: $\Delta \omega$; top-right: $\Delta \varpi$; bottom-right: $\Delta (\Omega+\varpi)$; $y$-axis: the numbers of libration events that spend the corresponding time in libration; $x$-axis: logarithmic time in the power of 10}
\label{stability-n}
\end{center}
\end{figure}

\begin{figure}[h]
\begin{center}
	\includegraphics[width=0.8\textwidth]{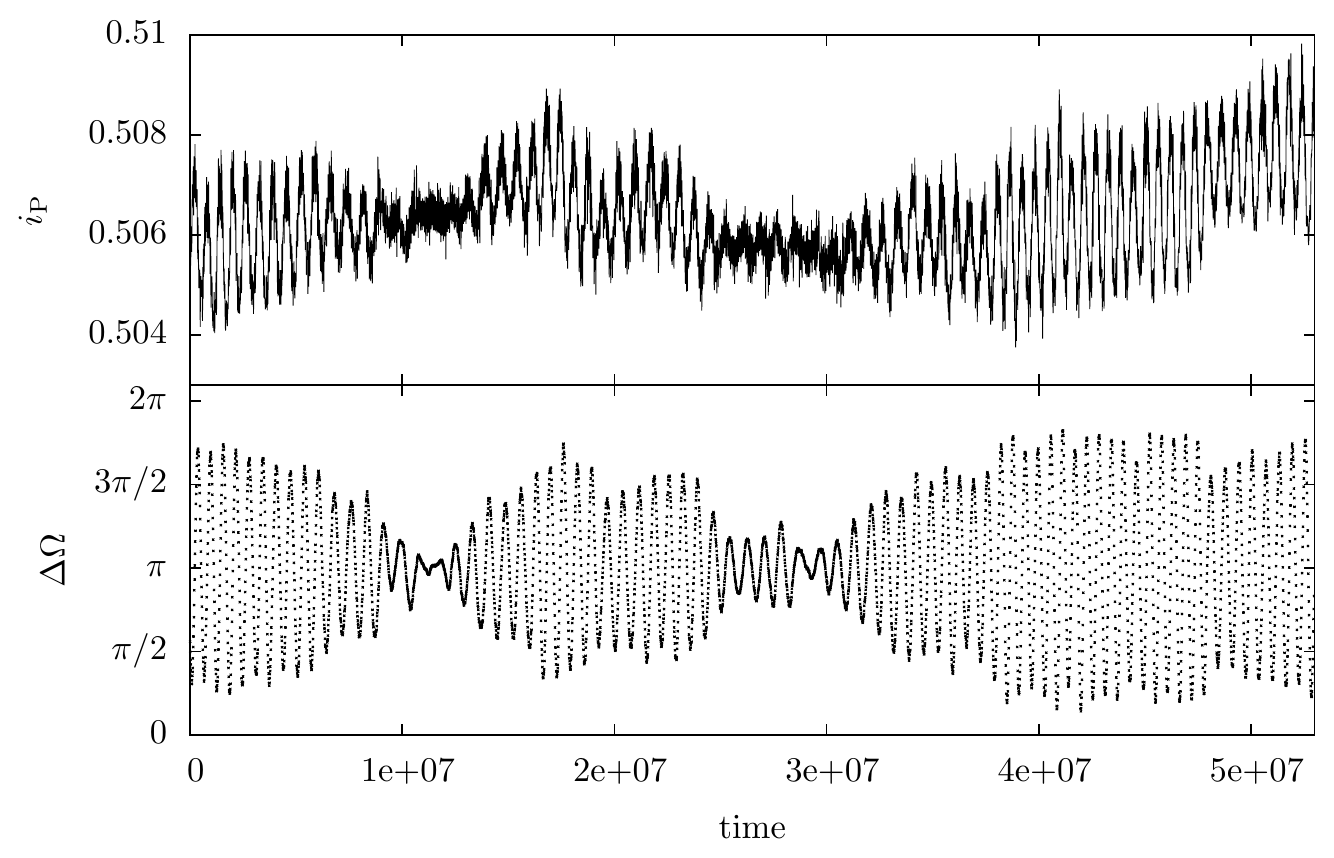}
\caption{A particle that stays in nodal libration for the entire integration duration of $5.3\times10^7$. Top panel: $i$; bottom panel: $\Delta \Omega$; $x$-axis: time}
\label{most-stable}
\end{center}
\end{figure}

In the $N$-body simulations, the evolution of eccentricity and inclination often experiences irregular short-period variations. It is then difficult to estimate the strengths of the resonances in terms of the amplitudes of the actions. Instead, we use the term ``stability'' to describe the resonances in real $N$-body simulations. We define stability as the longest duration of a single libration event from within the set of all such events. We extract the time duration for each libration event of the four resonances; we plot the distribution of durations of all such events in Fig. \ref{stability-n}. The nodal resonance is the most stable, as it can lock a test particle in libration for $\gg10^7$. Actually, we even have two particles remaining in libration for the entire integration of $\approx 5.3\times 10^7$; see Fig. \ref{most-stable}. The resonances in $\Delta \omega$ and $\Delta (\Omega+\varpi)$ are similarly stable; both can stay in libration for up to $\approx 10^7$. The resonance in $\Delta \varpi$ is the least stable: the libration cannot survive longer than $ \approx 5.6 \times 10^6$. These results correlate with the strength estimates in Subsect. \ref{sec-phase-space}: the resonance with the largest amplitude is the most stable.

\begin{figure}[h]
\begin{center}
	\includegraphics[width=0.8\textwidth]{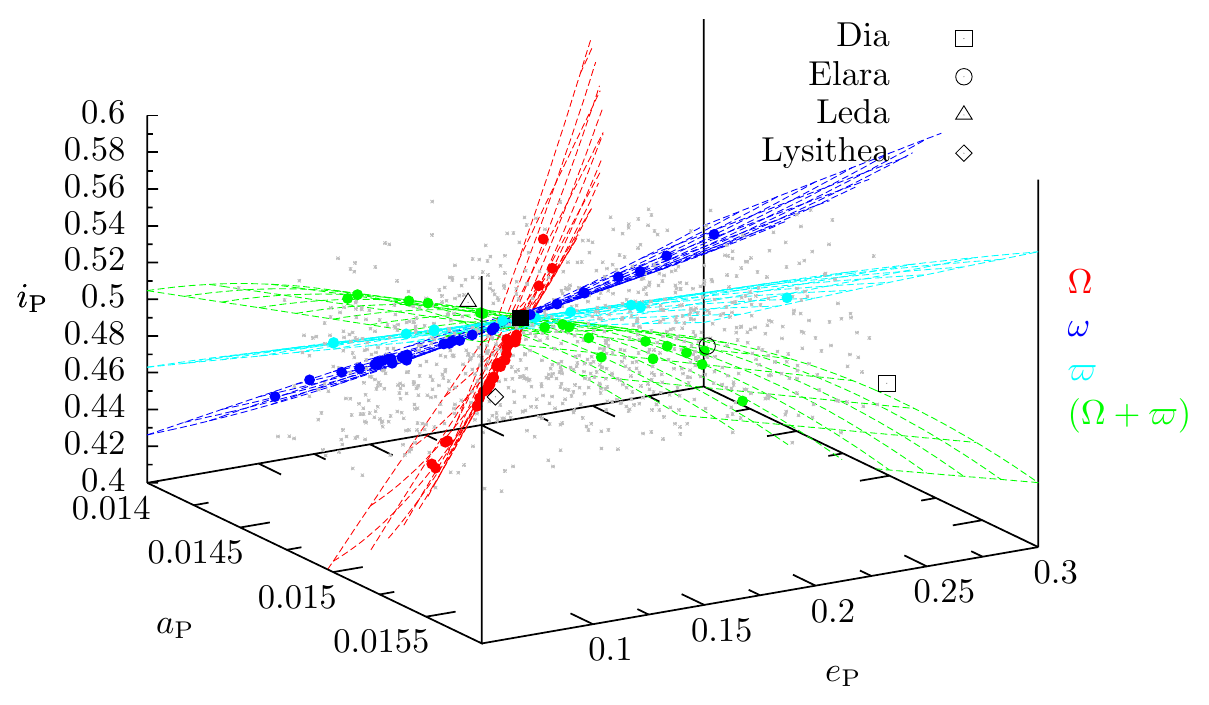}
\caption{The positions of the librators found in $N$-body simulations and the corresponding fitted SEPRs in $(a,e,i)$ space in comparison with all integrated particles and with the real family members. The big round points with different colours represent the librators of the four types. The surfaces with corresponding colours are the fitted SEPRs. The small grey points are the 1000 test particles. The unfilled square, circle, triangle and diamond represent Dia, Elara, Leda and Lysithea, respectively. The black square is Himalia, again}
\label{elements-resonance}
\end{center}
\end{figure}

We compute the mean orbital elements of a librator when inside resonance and plot them in Fig. \ref{elements-resonance}. We only use the averaged elements inside the libration, since as noted by \citet{Christou2005}, the elements may experience sudden changes in entering and/or leaving libration (e.g., Figs. \ref{n-hpw} and \ref{n-hh-change}). In the same figure, we show the following quadratic fit for each type of the resonances:
\begin{equation}
i= A_0+A_1 a +A_2 e+A_3 a^2+A_4 a e +A_5 e^2
\label{fit-surface}
\end{equation}
in which the $A_j$, $j=1,\ldots,5$, are the coefficients to be determined. These surfaces can be regarded as the SEPRs in the full problem since they originate from $N$-body simulations. Comparing them with the surfaces from semianalytical theory in Fig. \ref{equal-precession-all}, we find that the semianalytical model reproduces the relative positions and the overall slopes and shapes of the surfaces fairly well. The main reason for the differences between the two models for large inclination is that, the oscillation of eccentricity inside a Kozai cycle is not small here, contrary to our assumption of small variation of eccentricity in deriving Eq. \eqref{constant-kozai}. The fitted surface for $ \Omega$ has been used to plot the distance to the SEPR in Fig. \ref{n-hh} (top panel, dashed curve). As a reference, the real family members are shown as well. As expected, Lysithea is close to the SEPR for $\Omega$.

\begin{figure}[h]
\begin{center}
	\includegraphics[width=0.8\textwidth]{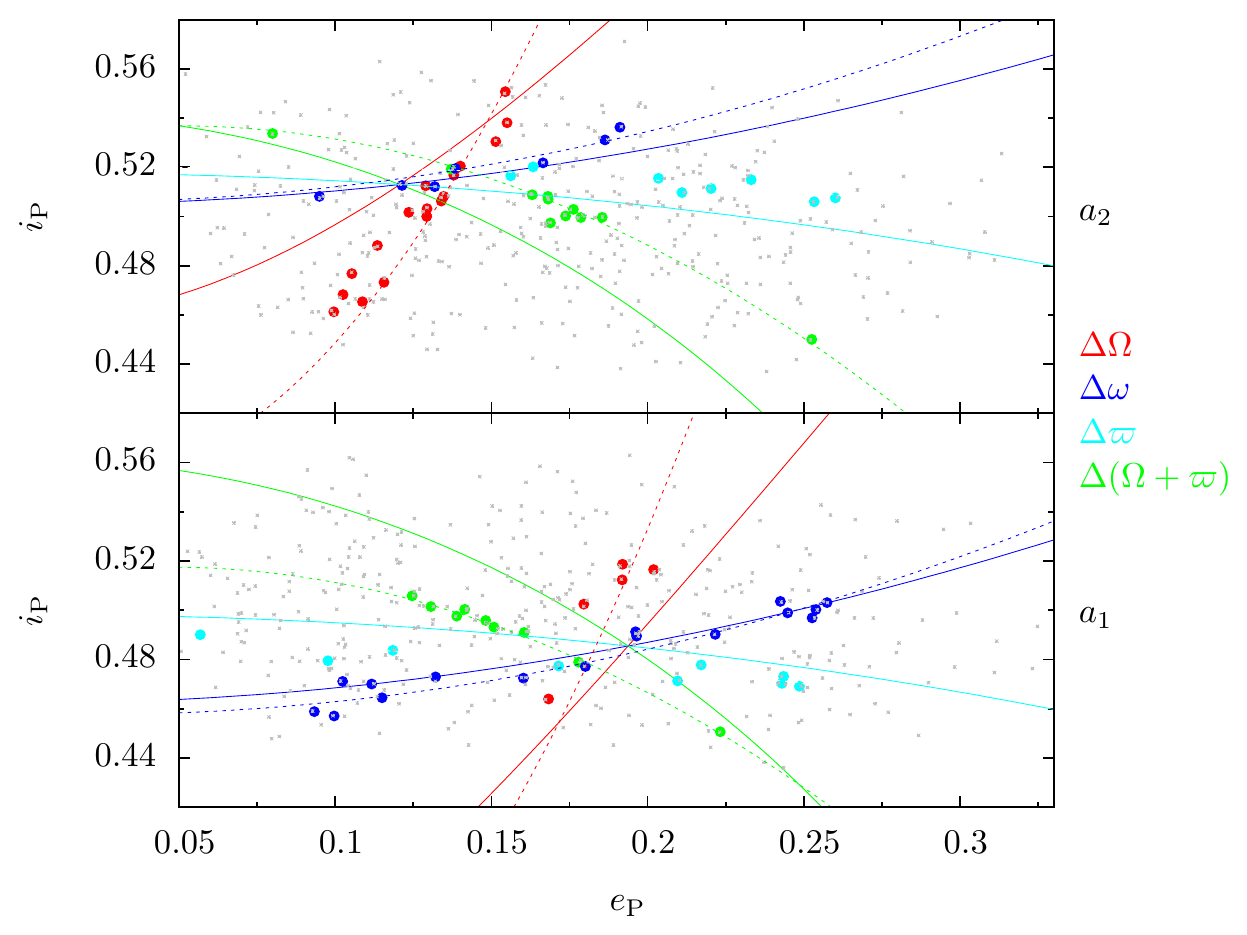}
\caption{The test particles that pass through the four types of resonance in the simulations with fixed initial $a$s and the corresponding curves of equal precession rate. Top panel: results with $a_\mathrm{P}=a_2$; bottom panel: results with $a_\mathrm{P}=a_1$; $y$-axis: $i$; $x$-axis: $e$. Plotted here is the mean eccentricity and inclination of the librators (of the four resonances and for the time slots in libration) as coloured round points. The solid coloured curves are from Kozai potential \eqref{constant-kozai} and the dashed curves are from the fit \eqref{fit-surface}. Small grey points are all the $2\times 400$ particles}
\label{fixed-a}
\end{center}
\end{figure}

\begin{figure}[h]
\begin{center}
	\includegraphics[width=0.8\textwidth]{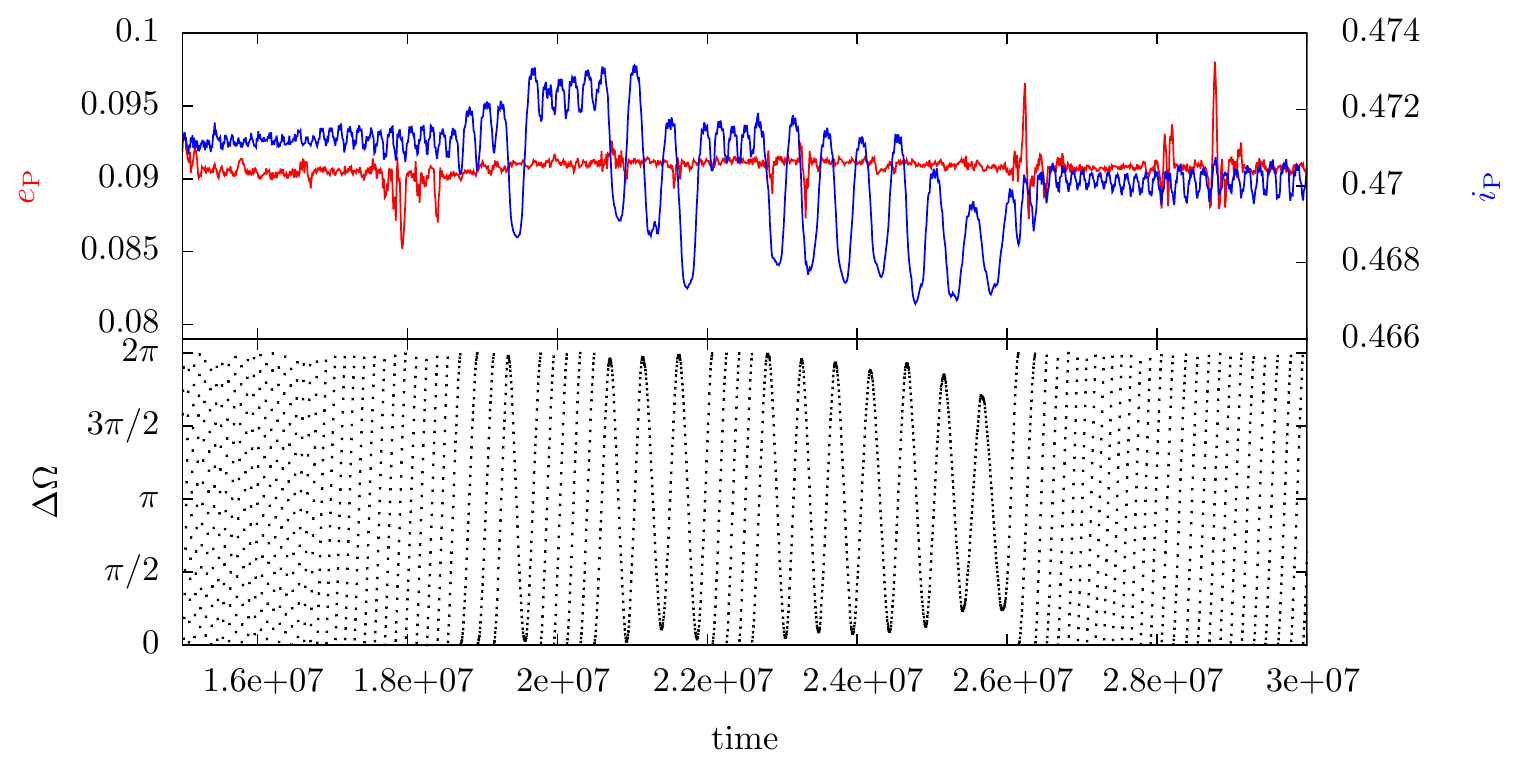}
\caption{An example of the change in inclination after passing through nodal resonance. This particular particle experiences a decrease of $\sim 0.001$ in inclination after the resonant passage}
\label{n-hh-change}
\end{center}
\end{figure}

To better highlight the positions of the resonant particles with respect to the SEPRs, we perform another set of simulations with $2\times 400$ test particles. The generation of initial conditions is, by and large, the same as the previous simulations except that here the initial osculating semimajor axes only assume one of the two values $a_1=0.98 $ and $a_2=1.02$ times the family mean semimajor axis. Then we plot the averaged $e$ and $i$ of the librators found in these $2\times400$ particles in the $(e,i)$ plane and compare them with the positions of the curves of equal precession rate (the intersections of the SEPRs with the $a=a_1$ and $a=a_2$ planes) in Fig. \ref{fixed-a}. Again, the curves of equal precession rate from semianalytical model (solid curves) agree with those from the fit (dashed curves). Note that, although the initial osculating semimajor axes are uniquely defined, they can vary significantly over the course of the simulations; thus the mean $a$ could be different from $a_1$ and $a_2$. This explains the excursion of a point from its corresponding fitted curve. For example, in the lower panel, the semimajor axis of the curves is exactly $a_1$ while for the points, the semimajor axes may be slightly different.

With the 1000 and $2\times 400$ test particles, we have 54, 52, 38 and 28 in resonances of $\Delta \Omega$,  $\Delta \omega$,  $\Delta (\Omega+\varpi)$ and  $\Delta \varpi$. They are all considered in fitting \eqref{fit-surface} while in Figs. \ref{stability-n} and \ref{elements-resonance}, only the 1000 particle integration results are plotted. We have visually checked these code detected librators and confirm that they are actual librators. However, since our method is insensitive to short duration libration events, more short-lived librators are likely.

As noted in \citet{Christou2005}, it was possible that the nodal resonant passage had a net effect in the sense that the inclination, after the libration passage, could change $\sim 0.1^\circ$, mostly increasing. In our simulations, we have confirmed the changes of similar amplitude; we find both directions are possible. Here, we show an example in Fig. \ref{n-hh-change}. These resonance-induced changes and the cumulative long-term consequences will be addressed in a forthcoming paper.

\section{Conclusion and discussion}
\label{sec-discussion}
We have investigated the dynamics of the Himalia irregular satellite group in a restricted four-body problem scheme where Jupiter, Sun and the satellite Himalia have finite masses while the test particle, representing a fictitious group member, is massless. We combine the Kozai with coorbital theories to construct a unified semianalytical model; the ``fast'' effects (operating on timescales comparable to that of a Kozai cycle) are removed. Then we arrive at a ``slow'' four-dimensional system. In this system, we are able to reproduce the main characteristics of a libration in $\Delta \Omega$ first observed in the numerical simulations in \citet{Christou2005}. We also identify two additional types of resonance involving the angles $\Delta \omega$ and $\Delta (\Omega + \varpi)$; their locations in phase space can be approximated by the surfaces of equal precession rate (SEPRs). In reducing the systems to two-dimensional ones, we determine the phase space structure for each of the three resonances. Using numerical integrations of the full Newtonian equations of motion, we confirm the results of the semianalytical model regarding the three types of resonance and find a fourth type involving $\Delta \varpi$. We analyse the occurrence rate and stability of these resonances in the $N$-body simulations. The libration periods of these four resonances are all of order $10^6$; specifically, the FFT-based mean period of nodal resonance is $5.73\times 10^5$ ($\sim$ 1.08 Myr). The main properties of the resonances are listed in Table \ref{resonances-summary}.

\begin{table}[h]
\caption{Summary of the properties of the four resonances studied in this work}
\label{resonances-summary}
\begin{center}
\begin{tabular}{ *7{>{\centering\arraybackslash}m{0.105\textwidth}} @{}m{0pt}@{}}
	\hline
	\hline
	\shortstack{libration \\ angle} & \shortstack{libration \\ centre} & \shortstack{conjugate \\ action} & \shortstack{conserved \\ quantity} & \shortstack{relation \\ of $e_\mathrm{P}$ and $i_\mathrm{P}$} & \shortstack{amplitude \\ of $e_\mathrm{P}$ and/or $i_\mathrm{P}$} & stability\\
	\hline
	$\Delta \Omega$ & $\pi$ & $i_\mathrm{P}$ & $e_\mathrm{P}$ & - & $\sim 10 ^{-3}$ & most stable\\
	\hline
	$\Delta \omega$ & $\pi$ & $e_\mathrm{P}, i_\mathrm{P}$ & ${( e_\mathrm{P}^2+i_\mathrm{P}^2)/2}$ & \shortstack{anti- \\ correlated} & $\sim 10 ^{-4}$ & \shortstack{moderately \\ stable}\\
	\hline
	$\Delta (\Omega+\varpi)$ & 0 & $e_\mathrm{P}, i_\mathrm{P}$ & ${( e_\mathrm{P}^2-i_\mathrm{P}^2)/2}$ & correlated & $\sim 10 ^{-4}$ & \shortstack{moderately \\ stable}\\
	\hline
	$\Delta \varpi$ & 0 & - & - & - & - & least stable\\
	\hline
\end{tabular}
\end{center}
\end{table}

In the semianalytical approach, the ``evection'' terms \citep[][see also Sect. \ref{sec-intro}]{Yokoyama2008} that relate the mean motion of the Sun with the apsidal rate of the satellite are omitted. The evection effect can be important for satellites far from the planet ($a \gtrsim  r_\mathrm{H}/2$; $r_\mathrm{H}$ is the Hill radius of the planet) where the precession rate is fast. In the planar prograde case, the evection potential to the lowest order in $\alpha$ has the form \citep{Frouard2010}
\begin{equation}
R_\mathrm{evec} \sim {{m_\odot \alpha^3}\over{ a_\odot}} e^2 \cos(2 \varpi -2 \lambda_\odot) ,
\end{equation}
where $\lambda_\odot$ is the Solar mean longitude. We have implemented the evection corrections as proposed in \cite{Cuk2004} and found that it does not improve the precision of our model significantly. To be exact, with the corrections, we are still unable to approximate the nodal rate accurately enough to locate the position of the SEPR (in the full $N$-body model). Thus the evection correction is dropped. In addition, though the octupole solar perturbation potential is zero owing to the circular orbit assumed for the Sun \citep[e.g.,][]{Krymolowski1999}, even higher order terms may contribute to the precession of the node and shift the location of resonance slightly. Regarding our simple way in removing $\omega$ in Eq. \eqref{quad_initial} and in reaching Eq. \eqref{constant-kozai}, there are semianalytical methods for this purpose \citep[][]{Henrard1990}. But that approach requires the introduction of new variables that have no apparent physical meaning, which we believe will complicate the interpretation of the results.

Our semianalytical model does not reproduce the apsidal resonance in $\Delta \varpi$. The original Kozai potential \eqref{quad_initial} is dependent on $e$ only to order 2 but on $i$ to higher orders (through $\cos i$) while in the coorbital potential (Eqs. \eqref{co-potential} and \eqref{co-potential-one}), $e$ and $i$ are symmetric. Thus this shortcoming of the semianalytical model results from the Kozai potential, specifically the truncated form of the solar forcing. The lack of higher order terms in $e$ causes the absence of $\bar G ^2$ in the final Kozai potential \eqref{min-kozai} while $\bar H ^2$ does exist. Thus after partial differentiation, the rate $\dot {\bar g}$ does not depend on its conjugate angular momentum $\bar G$ but $\dot {\bar h}$ does depend on $\bar H$. Hence we suggest that, the Kozai potential not only gives rise to the SEPRs, it must also play an important role in modulating $\dot {\bar h}$ (thus ${\mathrm{d} \Delta \Omega / \mathrm{d}t}$) through $\bar H$. Therefore, terms like $\bar G ^2$ in the solar perturbation is needed for the apsidal libration, which is only possible in higher order terms in $\alpha=a/a_\odot$. Indeed, if we artificially add a term $\bar G^2$ in the Kozai potential, the resonance in $\Delta \varpi$ can be retrieved in the semianalytical model. We believe this is why we do not observe apsidal libration in the semianalytical model while it does appear in the $N$-body simulations. This is probably why the apsidal resonance is the most unstable and the weakest (see Figs. \ref{n-ww} and \ref{stability-n}).

Although our semianalytical model is truncated in $e$ and $i$, we test it with real Himalia group inclinations ($\approx0.5$) and find that, the model does reproduce the three resonances at these inclinations. However, by dividing the real inclinations by a factor of 3, the $e$ and $i$ are similarly small and more consistent with the assumptions made in deriving the coorbital interaction potential \eqref{co-potential} \citep{Henon1986,Namouni1999}.

We have simplified the problem to a restricted four-body problem model. In reality, perturbations such as the oblateness of Jupiter and the effects of other planets exist. When these additional factors are considered, we suggest these resonances may still persist. In our analysis, the solar perturbation mainly arrests relative precession and makes the resonances possible, though terms like $\bar H^2$ do contribute directly to give rise to the resonances. On the timescales we are interested in, it has been simplified in the form \eqref{min-kozai} containing no angles. When a more elaborate model is constructed, probably similar simplifications can be made and small changes in the resonance locations (in the form of the SEPRs) will likely occur. In \citet{Christou2005}, Saturn and the jovian oblateness were taken into account and in such a configuration, he found the resonance in $\Delta \Omega$. Including other planets may introduce additional dynamical effects such as secular resonances \citep[see, e.g.,][]{Beauge2007,Frouard2011}.

Similar to high order mean motion resonance \citep{Moons1997} and secular resonance \citep{Bottke2001}, resonances involving other combinations of $\Delta \Omega$ and $\Delta \varpi$ may exist. They may be weaker and less stable, compared to the resonances described in this paper.

As shown in Figs. \ref{equal-precession-all} and \ref{elements-resonance}, the SEPRs are close together to each other, especially near the massive satellite. Consequently, the resonances may interact with each other and give rise to chaos; e.g., Fig. \ref{n-hpw}. Considering that Himalia may give other group members random kicks during close encounters, it is not strange that a test particle may pass through more than one type of resonance if we extend the $N$-body simulations long enough. Actually, in our 100 Myr integration, we observe an example transiting from libration in $\Delta \varpi$ to libration in $\Delta (\Omega+\varpi)$.

Considering that the retrograde irregular satellite groups are more prevalent in the solar system and that, they often have a much more massive member, the resonances explored in this paper for prograde satellites may have their counterparts in the retrograde case. For example, the Ananke and Carme groups of Jupiter have inclinations around $150^\circ$ and $165^\circ$, respectively. However, the most massive members in the two groups are only 0.004 and 0.02 of the mass of Himalia ($m_\mathrm{H}$) and this is possibly too small to allow resonance phenomena to exist. As an experiment, we have run $N$-body simulations to test if a prograde massive satellite of $0.01 m_\mathrm{H}$ can induce nodal resonance and the results is positive -- such low masses may work. As the cases of evection phenomenon (described in Sect. \ref{sec-intro}), the definition of $\varpi$ becomes the difference between $\Omega$ and $\omega$ for retrograde orbits; it is possible that the structure of the retrograde resonances is different, for instance, the libration centre may change. For the masses and orbital elements mentioned here, see footnote 1.

Recently, \citet{Novakovic2015} reported a nodal resonance similar to the one discussed here. Their resonance is due to Ceres and could be responsible for the inclination dispersion in the Hoffmeister asteroid family. In their case, the main perturbation comes from Jupiter and the mass ratio between Ceres and Jupiter is of order $10^{-7}$; in our situation, the mass ratio $m_\mathrm{H}/m_\odot$ is $10^{-12}$, arguably a more extreme case. Due to the similar nature of the two problems, we suggest that other types of resonance (e.g., the one involving $\Delta \omega$) may exist in the asteroid belt.

Finally, in Fig. \ref{elements-resonance} we see that, Dia, Elara and Leda are near the SEPR in $(\Omega+\varpi)$ while Lysithea is close to the surface in $\Omega$. To check whether the real satellites could be affected by the resonances, we perform an additional integration using the orbits of the real moons for 1000 Myr. We find that, Leda does enter the resonance in $\Delta (\Omega+\varpi)$. Thus some members of the actual family may have been affected by these resonances. This will be addressed in a future paper.


\begin{acknowledgements}
DL thanks Dr. Matjia \'{C}uk for correspondences on the evection phenomenon. We are grateful for the helpful comments from two anonymous referees, increasing the quality of the paper. The authors wish to acknowledge the SFI/HEA Irish Centre for High-End Computing (ICHEC) for the provision of computational facilities and support. Astronomical research at the Armagh Observatory is funded by the Northern Ireland Department of Culture, Arts and Leisure (DCAL). Fig. \ref{model-illustration} is produced using LibreOffice Draw; all the other figures are generated with gnuplot.
\end{acknowledgements}


\begin{appendices}
\section{Derivation of the ``averaged'' Kozai Hamiltonian}
\label{sec-kozai-derivation}
We derive the Kozai Hamiltonian \eqref{constant-kozai} from its original form \eqref{quad_initial} here. First, we substitute $e$ and $i$ in Eq. \eqref{quad_initial} in terms of the angular momenta $G$ and $H$. Then we solve for $G$ so $G=G(F,H,\omega,const.)$, where $const.$ refers to combinations of $k$, $m_\odot$, $a$ and $a_\odot$. Note that in this expression, $F$, $H$ and $const.$ are all constant and only $\omega$ and $G$ are variable. Furthermore, it can be shown with this equation that the variation in eccentricity during a Kozai cycle, $\Delta e^2$ is a small quantity (e.g., for Himalia's nominal orbit, $\Delta e^2\lesssim 0.03$). Therefore, we assume that the angular momentum $G\propto\sqrt{1-e^2}$ can be expressed as
\begin{equation}
G^2=b_0+b_1 \cos 2  \omega + \dots ,
\end{equation}
where the coefficients $b_1\ll b_0$. We can then express $b_0$ and $b_1$ as functions of $F$, $H$ and $const.$ by comparing the above equation with $G=G(F,H,\omega,const.)$. Now, if we further omit the term $b_1 \cos 2  \omega$ we have
\begin{equation}
\label{to-solve-F}
G^2=b_0=b_0(F,H,const.)\propto{1-\langle e\rangle^2}\,.
\end{equation}
Here, $\langle e\rangle$ is not the $e$ in the original Hamiltonian \eqref{quad_initial} but the averaged $e$ over a Kozai cycle. Meanwhile, we also have $H^2\propto(1-\langle e\rangle^2)\cos^2\langle i\rangle$ and $\langle i\rangle$ is the averaged $i$. Then we solve Eq. \eqref{to-solve-F} for $F$. So now $F$ is a function of $H$, $b_0=G^2$ and $const.$. After substituting $\langle e\rangle$ and $\langle i\rangle$ into it, we arrive at the ``averaged'' Kozai Hamiltonian \eqref{constant-kozai}. The new eccentricity and inclination are constant under the solar forcing; for simplicity, we still use the old notations $e$ and $i$.

\section{The non-canonical property of coorbital potential}
\label{sec-non-canonical}
As stated in Subsect. \ref{sec-analytical}, the coorbital potential \eqref{co-potential} (and hence \eqref{slow-coorbital}) is not canonical in the variables used in Eq. \eqref{slow-kozai}. To prove this, we presume it is canonical and introduce the following canonical transformations. For the first transformation, we name the variables on the left side as ``absolute Cartesian'' variables.
\begin{equation}
\begin{aligned}
\bar{G}_\mathrm{y,H}&=& \sqrt{2 \bar G_\mathrm{H}}  \sin{\bar{g}_\mathrm{H}} ,\quad \bar{G}_\mathrm{x,H}&=& \sqrt{2 \bar G_\mathrm{H}}  \cos{\bar{g}_\mathrm{H}}  ;
\\
 \bar{H}_\mathrm{y,H}&=& \sqrt{2 \bar H_\mathrm{H}}  \sin{\bar{h}_\mathrm{H}} ,\quad \bar{H}_\mathrm{x,H}&=& \sqrt{2 \bar H_\mathrm{H}}  \cos{\bar{h}_\mathrm{H}}  ;
\\
\bar{G}_\mathrm{y,P}&=& \sqrt{2 \bar G_\mathrm{P}}  \sin{\bar{g}_\mathrm{P}} ,\quad \bar{G}_\mathrm{x,P}&=& \sqrt{2 \bar G_\mathrm{P}}  \cos{\bar{g}_\mathrm{P}}  ;
\\
 \bar{H}_\mathrm{y,P}&=& \sqrt{2 \bar H_\mathrm{P}}  \sin{\bar{h}_\mathrm{P}} ,\quad  \bar{H}_\mathrm{x,P}&=& \sqrt{2 \bar H_\mathrm{P}}  \cos{\bar{h}_\mathrm{P}} .
\end{aligned}
\end{equation}
On the right-hand sides of the above expressions, $(\bar g,\,\bar G)$ and $(\bar h,\,\bar H)$ are conjugate Poincar\'{e} variables with subscripts ``H'' and ``P'' referring to Himalia and a particle, respectively. Then another transformation is introduced. The variables on the left are referred to as ``relative Cartesian'' variables.
\begin{equation}
\label{relative-xy}
\begin{aligned}
\bar{G}_\mathrm{y,t}&=&\bar{G}_\mathrm{y,P}/\sqrt{2}+\bar{G}_\mathrm{y,H}/\sqrt{2} ,\quad  \bar{G}_\mathrm{x,t}&=&\bar{G}_\mathrm{x,P}/\sqrt{2}+\bar{G}_\mathrm{x,H}/\sqrt{2} ;
\\
\bar{H}_\mathrm{y,t}&=&\bar{H}_\mathrm{y,P}/\sqrt{2}+\bar{H}_\mathrm{y,H}/\sqrt{2}  ,\quad  \bar{H}_\mathrm{x,t}&=&\bar{H}_\mathrm{x,P}/\sqrt{2}+\bar{H}_\mathrm{x,H}/\sqrt{2} ;
\\
\bar{G}_\mathrm{y,r}&=&\bar{G}_\mathrm{y,P}/\sqrt{2}-\bar{G}_\mathrm{y,H}/\sqrt{2} ,\quad \bar{G}_\mathrm{x,r}&=&\bar{G}_\mathrm{x,P}/\sqrt{2}-\bar{G}_\mathrm{x,H}/\sqrt{2}  ;
\\
 \bar{H}_\mathrm{y,r}&=&\bar{H}_\mathrm{y,P}/\sqrt{2}-\bar{H}_\mathrm{y,H}/\sqrt{2} ,\quad \bar{H}_\mathrm{x,r}&=&\bar{H}_\mathrm{x,P}/\sqrt{2}-\bar{H}_\mathrm{x,H}/\sqrt{2}  .
\end{aligned}
\end{equation}
This relates the variables with subscript ``r'' to the variables used in the coorbital theory as given by \citet{Namouni1999}; see the paragraph containing Eqs. \eqref{co-potential}-\eqref{co-euqation-motion}.

Now note that, since we assume the coorbital potential is canonical in the original Poincar\'{e} variables, it should preserve the Hamiltonian property in the variables above. However, since the coorbital potential contains only the relative elements, i.e., in the relative Cartesian variables, it only depends on the variables with subscript ``r''. Hence, if we consider the Jupiter-Himalia-particle system. The quantities with subscript ``t'' will be conserved. However, this is not possible. Taking $\bar{G}_\mathrm{y,t}=\bar{G}_\mathrm{y,P}/\sqrt{2}+\bar{G}_\mathrm{y,H}/\sqrt{2}$ for instance. Since the test particle is massless, it should have no influence on Himalia and thus the second term on the right hand side $\bar{G}_\mathrm{y,H}/\sqrt{2}$ is constant. From the above reasoning, we know that, $\bar{G}_\mathrm{y,t}$ should be conserved. This means that, the first term on the right hand side, $\bar{G}_\mathrm{y,P}/\sqrt{2}=\sqrt{\bar G_\mathrm{P}}  \sin{\bar{g}_\mathrm{P}}$ will no change. Similarly, $\sqrt{\bar G_\mathrm{P}}  \cos{\bar{g}_\mathrm{P}}$, $\sqrt{\bar H_\mathrm{P}}  \sin{\bar{h}_\mathrm{P}}$ and $\sqrt{\bar H_\mathrm{P}}  \cos{\bar{h}_\mathrm{P}}$ should all be constant. If so, the orbital elements of the particle are not evolving under the perturbation of Himalia. Of course this is not true -- the coorbital potential is not a Hamiltonian in the variables used here.

However, the level-curve like structure in the phase diagrams Figs. \ref{phase-nodal}, \ref{phase-wmh} and \ref{phase-hpw} of the three types of resonance suggests the existence of a conserved quantity (a Hamiltonian) for these two-dimensional systems. We have tried to construct one of the form of $F_\mathrm{T}=F_\mathrm{S}+A \times R_\mathrm{S}$ (see Eqs. \eqref{slow-kozai} and \eqref{slow-coorbital}), where $A$ is a constant depending on some non-changing  parameters. For instance, for the case of nodal resonance, $A$ could be a function of the masses, the semimajor axes and the eccentricities (but not a function of inclinations). However, the solution for $A$ seems to be nonexistent (at least for the nodal resonance). Thus we argue that, the conserved quantity of the two-dimensional systems might be of more complicated forms.

\end{appendices}



\end{document}